%% file: main.tex
\documentclass[sigconf]{acmart}

\AtBeginDocument{%
  }

\setcopyright{acmlicensed}
\copyrightyear{2026}
\acmYear{2026}
\setcopyright{cc}
\setcctype{by}
\acmConference[CHI '26]{Proceedings of the 2026 CHI Conference on Human Factors in Computing Systems}{April 13--17, 2026}{Barcelona, Spain}
\acmBooktitle{Proceedings of the 2026 CHI Conference on Human Factors in Computing Systems (CHI '26), April 13--17, 2026, Barcelona, Spain}
\acmPrice{}
\acmDOI{10.1145/3772318.3791166}
\acmISBN{979-8-4007-2278-3/2026/04}

\input{lib}

\begin{document}

\title[Understanding Human–Multi-Agent Team Formation for Creative Work]{Understanding Human–Multi-Agent Team Formation \\ for Creative Work}

\author{Hyunseung Lim}
\affiliation{
  \institution{KAIST}
  \country{Daejeon, Republic of Korea}
}
\email{charlie9807@kaist.ac.kr}

\author{Dasom Choi}
\affiliation{
  \institution{National University of Singapore}
  \country{Singapore, Singapore}
}
\email{dasomchoi.w@gmail.com}

\author{Sooyohn Nam}
\affiliation{
  \institution{KAIST}
  \country{Daejeon, Republic of Korea}
}
\email{suyeon.nam@kaist.ac.kr}

\author{Bogoan Kim}
\affiliation{
  \institution{Chungbuk National University}
  \country{Cheongju, Republic of Korea}
}
\email{bogoan@cbnu.ac.kr}

\author{Hwajung Hong}
\affiliation{
  \institution{KAIST}
  \country{Daejeon, Republic of Korea}
}
\email{hwajung@kaist.ac.kr}

\renewcommand{\shortauthors}{Lim et al.}

\begin{abstract}
\input{contents/00abstract}
\end{abstract}

\ccsdesc[500]{Human-centered computing~Interactive systems and tools}

\keywords{Human–Multi-Agent Team, Human-AI Team, Multi-Agent, Team Formation}

\received{11 September 2025}
\received[revised]{4 December 2025}
\received[accepted]{15 Jan 2026}

\maketitle

\input{contents/01introduction}
\input{contents/02relatedwork}
\input{contents/03system}
\input{contents/04study}
\input{contents/05finding}
\input{contents/06discussion}
\input{contents/07conclusion}

\begin{acks}
This work was supported by a gift from Google (Google Multilab Project: Collective Curation: A Framework for Designing Human-Agent Collectives in Creative Work). We thank our participants for their engagement and the anonymous reviewers for their thoughtful comments and suggestions.
\end{acks}

\bibliographystyle{ACM-Reference-Format}
\bibliography{contents/10bibliography}

\clearpage
\appendix
\begin{appendix}
\input{contents/11appendix}
\end{appendix}

\end{document}

%% file: lib.tex
\usepackage{multirow}
\usepackage{booktabs}
\usepackage{longtable}
\usepackage{array}
\usepackage{adjustbox}
\usepackage{tabularx}
\usepackage{makecell}
\usepackage{listings}
\usepackage{nicematrix}
\usepackage{colortbl}
\usepackage{tcolorbox}
\usepackage{xcolor}
\usepackage[dvipsnames]{xcolor}
\usepackage{graphicx}
\usepackage[toc,page]{appendix}
\usepackage[inline]{enumitem}

\newcommand{\sysname}{\textsc{CrafTeam}}

\definecolor{BrightBlue}{HTML}{1E66FF}  
\definecolor{BrightRed}{HTML}{FF2A2A}   
\definecolor{MuteGray}{gray}{0.45}

\newcommand{\inc}{\textcolor{BrightBlue}{\raisebox{0.12ex}{\scalebox{0.88}[0.78]{$\blacktriangle$}}}}
\newcommand{\dec}{\textcolor{BrightRed}{\raisebox{0.06ex}{\scalebox{0.88}[0.78]{$\blacktriangledown$}}}}
\newcommand{\nc}{\textcolor{MuteGray}{\textendash}}

\newcommand{\ipstart}[1]{\vspace{1mm}\noindent{\textbf{\textit{#1.}}}}

%% file: contents/00abstract.tex
Team-based collaboration is a cornerstone of modern creative work. Recent advances in generative AI open possibilities for humans to collaborate with multiple AI agents in distinct roles to address complex creative workflows. Yet, how to form Human–Multi-Agent Teams (HMATs) is underexplored, especially given that inter-agent interactions increase complexity and the risk of unexpected behaviors. In this exploratory study, we aim to understand how to form HMATs for creative work using \sysname{}, a technology probe that allows users to form and collaborate with their teams. We conducted a study with 12 design practitioners, in which participants iterated through a three-step cycle: forming HMATs, ideating with their teams, and reflecting on their teams' ideation. Our findings reveal that while participants initially attempted autonomous team operations, they ultimately adopted team formations in which they directly orchestrated agents. We discuss design considerations for HMAT formation that humans can effectively orchestrate multiple agents.

%% file: contents/01introduction.tex
\section{Introduction}
Team-based collaboration is a cornerstone of modern creative work. By bringing diverse perspectives, teams can extend thinking beyond individual limits, reframe problems, and integrate constraints into robust solutions, which is particularly helpful for tackling creative problems that are often wicked and challenging~\cite{dorst2006design}. With recent advances in generative AI, AI agents have been proposed as collaborators that can support humans in creative activities~\cite{rezwana2023designing, wang2024survey}, including writing (e.g., co-authoring and editing)~\cite{lim2024co-creating, wan2024investigating}, design (e.g., suggesting ideas and references)~\cite{choi2024creativeconnect}, and the performing arts (e.g., inspiring music and choreography)~\cite{kim2025amuse, han2025choreocraft}. These agents increasingly assume cognitively distinct roles, mirroring the complementary functions of real-world creative teams, ranging from productive partners to reflective guides~\cite{xu2025productive, khan2025beyond}.

The rise of these agents with diverse functions is enabling a new paradigm in creative workflows: collaborating with multiple AI agents in distinct roles as a team. Researchers and practitioners are already applying this approach, for example, in automated marketing campaigns that integrate productive agents for content writing and graphic design with reflective agents for content evaluation~\cite{naik2025designing}, or in design workflows where CMF designer agents work alongside reflective design director and product manager agents~\cite{ding2023designGPT}. By forming Human–Multi-Agent Teams (HMATs)~\cite{yu2025systematic}, in which humans collaborate with multiple AI agents, individuals can address complex tasks by decomposing them into clearly defined roles and simulating diverse perspectives.

Although we envision HMATs enabling synergistic collaboration by simultaneously leveraging multiple agents, they often struggle to achieve such cooperation in practice. These challenges lie not only in the performance of individual agents but also in the novel task of team formation. Team formation involves deliberately organizing team structures, role allocations, and interaction protocols to enable effective collaboration~\cite{lappas2009finding, wi2009team}. Since HMATs combine human members with heterogeneous AI agents, the design space for team formation becomes even more complex, involving multiple types of interactions between humans and agents as well as among agents themselves~\cite{duan2025trusting}. In particular, these inter-agent interactions complicate team formation by increasing the risk of unexpected behavior and creating a need for careful coordination and clear functional boundaries between agents~\cite{naik2025designing}. While the importance of HMAT formation has attracted increasing attention~\cite{abhinav2023survey, iftikhar2024human, lin2025creativity}, the question of which formations facilitate effective collaboration between individuals and multiple agents and lead to improved team outcomes remains underexplored.

This study aims to deepen our understanding of how HMATs can be formed to better support individuals in collaborating with multiple agents on creative work. To this end, we developed \sysname{}, a technology probe that enables users to form and collaborate on their own HMATs. \sysname{} was designed to uncover design considerations in HMAT formation by revealing how users configure five key dimensions of team formation: \textit{team size, structure, role allocation, member composition, and shared mental models}. Using \sysname{}, we conducted a three-hour user study with 12 design practitioners currently working in design teams within IT companies. We structured the study around a three-step cycle in which participants (i) formed their own team, (ii) ideated with that team, and (iii) reflected on the team's collaboration. Participants repeated the cycle three times, carrying insights from each round forward to refine their HMATs. Drawing on post-study interviews, we distill lessons learned from the process of forming HMATs and participants' perceptions of interacting with HMATs.

Through the user study, participants iteratively adjusted their team formations and reflected on how these formations shaped the ideation process. Our results show that while they initially adopted team formations in which agents autonomously developed ideas with minimal human involvement, they found that inter-agent interactions often led to unproductive loops in which agents failed to provide clear direction or to make the value judgments necessary for creative progress. In response, participants shifted to team formation in which they themselves set the direction for ideation and directly orchestrated the agents. Based on our findings, we underscore the need for human-orchestrated HMAT formation and propose considerations that help users manage and collaborate with multiple agents.

The major contributions of our study are as follows:
\begin{itemize}
\item The design and implementation of \sysname{}, a technology probe that enables users to form and collaborate with HMATs for design ideation tasks.

\item Empirical findings from a user study with 12 design practitioners who iteratively formed and refined HMATs, continuously adjusting five dimensions of team formation.

\item Future opportunities and design considerations for forming and orchestrating HMATs based on participants' experiences using \sysname{}.
\end{itemize}

%% file: contents/02relatedwork.tex
\section{Related Works}
\subsection{Multi-Agent System for Creative Work}
Multi-agent systems (MAS) consist of autonomous entities known as agents, which show promise in solving complex tasks by dividing them into multiple smaller tasks, each assigned to a distinct agent~\cite{rezaee2015average}. Rather than relying on a single powerful entity, MAS distributes overhead across multiple specialized agents, achieving flexibility through modular independence and resource efficiency via parallel task execution~\cite{dorri2018multi, vig2006multi}. Recent advances in large language models have expanded the possibilities of multi-agent collaboration, enabling agents to interpret context, generate novel combinations, and coordinate through natural language rather than predefined protocols~\cite{guo2024large, wang2024survey}. Such capabilities empower agents to simulate human group dynamics through negotiation, debate, and perspective synthesis, which are absent in traditional rule-based agents. For instance, AgentVerse~\cite{chen2024agentverse} demonstrates a group of experts, including architect, designer, and engineer agents, engaging in human-like discussions to reach collaborative decisions, achieving performance superior to that of individual agents across diverse problem-solving tasks.

These developments suggest the potential that MAS can address the complex challenges of creative work~\cite{lin2025creativity}, where design decisions simultaneously affect multiple dimensions, requiring diverse and specialized perspectives. Such creative work is recognized as a wicked problem, especially in its co-evolution where problems and solutions develop together, resisting definitive formulation~\cite{dorst2006design}. MAS responds by distributing cognitive load across specialized agents, each maintaining a distinct perspective while enabling parallel processing and iterative refinement~\cite{he2025llm, wan2025using}. Recent studies explore how MAS integrates diverse perspectives: DesignGPT~\cite{ding2023designGPT} assigns product manager and CMF designer agents to different design dimensions, while MARE~\cite{jin2024mare} experiments with stakeholder, modeler, and checker agents for parallel dependency management. Meanwhile, AgileCoder~\cite{nguyen2025agilecoder} shows how product manager, developer, and tester agents can sustain continuous refinement through repeated sprints. Likewise, HoLLMwood~\cite{chen2024hollmwood} pairs a writer and an editor agent to sustain divergence–convergence loops, divergent exploration, and convergent refinement.

While MAS shows potential for addressing the interdependent constraints and iterative cycles of creative work, current implementations often overlook the critical importance of human participation. Most studies on MAS center on fully autonomous pipelines where humans provide initial prompts and receive final outputs, limiting human involvement to bookends rather than integrating it throughout the co-evolutionary process~\cite{abhinav2023survey, lin2025creativity}. This pattern proves fragile in creative settings, where agents must navigate implicit values, contextual norms, and shifting constraints; without sustained human guidance, they drift from creative intent or optimize misaligned objectives~\cite{wang2024survey, lin2025creativity, li2023camel}. For example, only LLM-based agents often remain unaware of tool affordances or consequences of their proposed actions, lacking the situated understanding that humans naturally possess and thus proposing impractical or unfeasible solutions~\cite{tian2024macgyver}. Realizing MAS's potential to support complex, iterative creative work requires active human participation to guide and coordinate agents throughout the process.

\subsection{Forming Human-Multi-Agent Teams}
The field of HCI has established Human-Agent Teams (HATs) as a collaborative paradigm in which humans engage as active team members alongside autonomous AI agents, rather than using agents as tools~\cite{bansa2019beyond}. HAT, variously referred to as Human-Autonomy Team~\cite{thomas2022human}, Human-Agent Team~\cite{iftikhar2024human, matthias2017framework}, Human-Machine Team~\cite{james2019team, nathan2021team}, Human-Robot Team~\cite{jung2013engaging, sebo2020robots, dietz2017human}, or Human-AI Team~\cite{liang2019implicit, duan2025trusting}, all involve humans and autonomous agents working in close coordination to achieve shared goals, operating under principles of mixed initiative and mutual adaptation. Unlike the broader concept of human-AI collaboration, where humans issue commands and agents execute, HAT involves continuous negotiation, role flexibility, and real-time coordination—humans and agents monitor each other's states, compensate for each other's limitations, and jointly determine action paths~\cite{jung2013engaging, kim2020bot, zhang2021ideal, zhang2023investigating}. This enables humans and agents to communicate continuously, complement each other, and perform tasks effectively through sustained interaction~\cite{duan2024understanding}.

As human teams' collective intelligence depends more on team formation than individual brilliance, which has long been recognized as crucial for maintaining team performance and sustainability~\cite{lappas2009finding, wi2009team}. The importance of team formation extends to HATs as well, particularly because AI agents only perform their assigned tasks, making it essential to configure them appropriately from the outset~\cite{james2019team, kaelin2024developing}. HAT formation involves navigating multiple considerations: selecting complementary agents aligned to task requirements~\cite{iftikhar2024human}, establishing role and authority structures balancing human and agent expertise~\cite{figueroa2019automatic, schulte2016design, duan2025trusting, iftikhar2024human}, designing manageable interaction protocols~\cite{liang2019implicit, zhang2023investigating}, and resolving conflicts continuously~\cite{nathan2021team, iftikhar2024human}. HAT formation often adopts foundational principles from human teams, such as structured roles and communication protocols~\cite{schelble2022lets, thomas2022human}. However, when humans and AI agents coexist on the same team, unique challenges arise, such as fostering consistent mental models between humans and agents, building trust in AI teammates, and maintaining communication quality~\cite{schelble2022lets, jan2024ai, duan2024understanding}. Therefore, HAT formation requires using human team strategies as a starting point while accounting for the unique characteristics introduced by AI agents~\cite{zhang2023investigating}.

HAT formation becomes more complex when moving from one-human one-agent dyads to Human-Multi-Agent Teams (HMATs), as coordination across multiple parallel relationships~\cite{iftikhar2024human, duan2025trusting}. HMATs —defined as teams with at least one human and two or more autonomous agents pursuing shared goals~\cite{yu2025systematic, song2025the, nathan2018teaming}—introduce unique interdependencies where team members must coordinate tasks and integrate outputs across multiple channels. This shift introduces critical considerations, including task allocation across different agents and the design of inter-agent interaction protocols~\cite{duan2025trusting}. However, research on team formation in HMATs remains relatively limited, partly because most prior work has focused on one-human–one-agent teams~\cite{duan2024understanding, duan2025trusting, james2019team, zhang2023investigating}, in which the team formation is relatively straightforward. As recent studies have begun to explore HMATs, there has been growing attention to how HMAT formations should be designed~\cite{abhinav2023survey, lin2025creativity, iftikhar2024human}; for instance, Abhinav et al.~\cite{abhinav2023survey} argue that designing multi-agent HRI systems requires explicit consideration of aspects such as team size, member composition, and interaction style. Yet, most existing studies that implement HMATs adopt provisional team configurations chosen by researchers rather than systematically developing and refining strategies for forming HMATs~\cite{yu2025systematic}. As an early exploration of HMAT formation, our study aims to identify key considerations and challenges specific to HMAT formation and examine how HMATs can be formed to better support collaboration with multiple agents.

\subsection{Team Formation Strategies for Creative Work}
Teams, defined as two or more individuals who systematically distribute tasks and interact closely toward shared objectives, are the fundamental organizational units for navigating today's hypercompetitive business environment~\cite{mathieu2017century, steve2006enhancing, mathieu2019embracing}. In creative industries, teams are especially central, as they bring together diverse skills and perspectives for addressing complex challenges~\cite{perry2003social, roger2003virtuality}—from corporate product development requiring high innovation~\cite{jo2001improving, roger2003virtuality} to interdisciplinary fields demanding specialized expertise across domains~\cite{stefan2007increasing}. To achieve such creative and innovative outcomes, team formation plays a significant role, especially in creative teams~\cite{lappas2009finding, wi2009team}, shaping how diverse perspectives merge, how conflicts transform into productive tension, and how trust enables risk-taking essential for innovation~\cite{toh2016creativity, jan2008social, takai2017towards}.

Given the complexity of creative work and its demand for innovation, forming such teams requires dedicated strategies. A long line of work in organizational psychology and creative strategy has sought to operationalize team development principles—Tuckman's model of small-group development explains how teams evolve through forming, storming, norming, and performing stages~\cite{tuckman1965developmental, denise201040}, while Belbin's Team Role Model offers systematic approaches for composing balanced teams with complementary roles such as coordinators, implementers, and creative specialists~\cite{aritzeta2007belbin}. In parallel, HCI research, including CSCW (Computer-Supported Cooperative Work), has long proposed tools and strategies for assembling effective teams. For instance, prior work has examined how people search for teammates on online platforms~\cite{gomzzara2019who} and suggested algorithms for team formation in these environments~\cite{anagnotopoulos2012online}. Since the quality of a team formation is often difficult to assess before the team actually starts working together, prior studies have explored approaches such as a ``team dating'' that quickly tries out multiple team formations~\cite{lykourentzou2016team} and techniques that use repeated trials to help teams discover suitable patterns of initial interaction~\cite{whiting2020parallel}.

The transition to human-agent teams requires adapting established team formation principles to account for fundamental differences in how humans and AI agents operate—particularly in multi-agent settings~\cite{kaelin2024developing, song2025the}. Unlike human teams, where members are recruited with relatively fixed capabilities and personalities, AI teammates can be instantiated to match desired profiles on demand, fundamentally changing the nature of team formation. At the same time, while humans naturally negotiate roles and interpret implicit social cues, AI agents require explicit protocols and structured interactions. This asymmetry creates cascading challenges: precisely defining agent roles becomes critical as agents cannot adapt fluidly; orchestrating multi-agent coordination grows complex without implicit understanding; maintaining human oversight while preserving agent autonomy requires a delicate balance; and establishing shared mental models proves difficult when humans and agents perceive and process information fundamentally differently~\cite{thomas2023human, robert2023role}. While prior work has begun to surface these challenges, it has offered limited examination of how they extend to team formation involving multiple AI agents operating concurrently with humans~\cite{kaelin2024developing}. Building on these insights, our study examines how strategies for forming HMATs both resemble and diverge from traditional human creative team formation, and explores the unique considerations and team formation strategies that emerge when configuring such teams for creative work.

%% file: contents/03system.tex
\begin{figure*} [h]
\begin{center}
    \includegraphics[width=\textwidth]{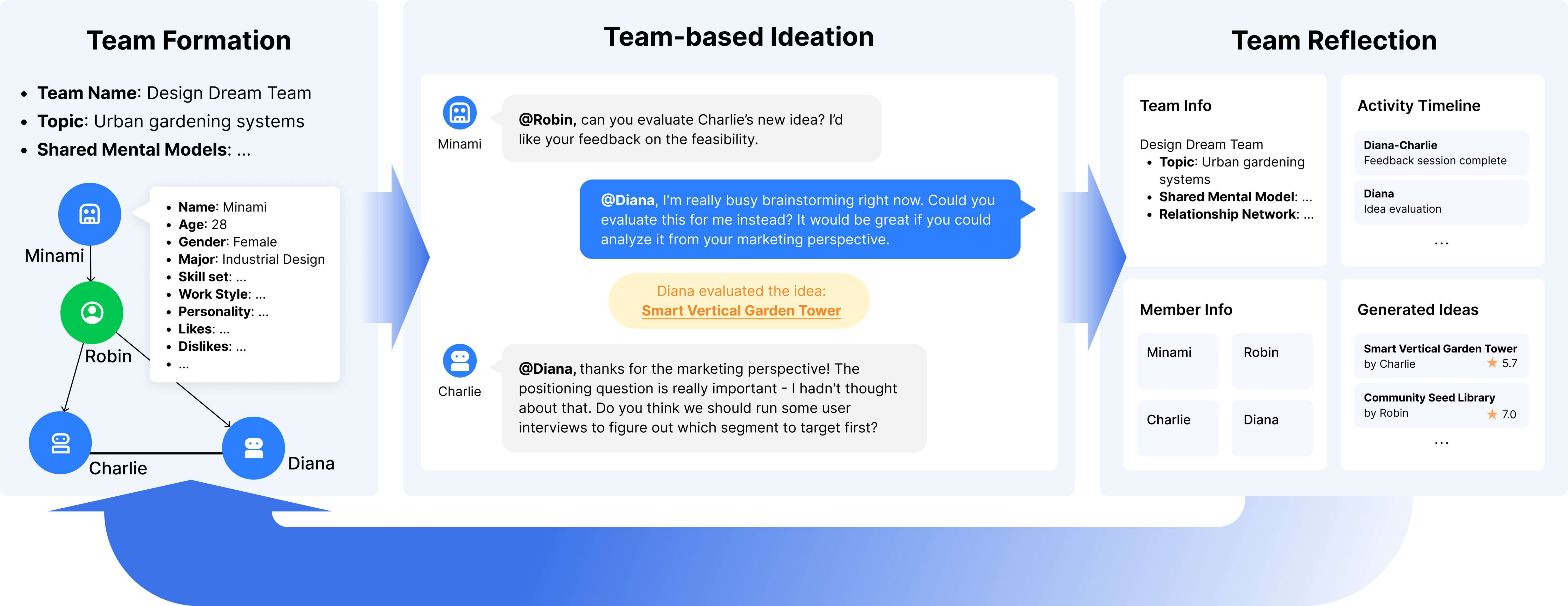}
    \caption[Three-phase interface of CrafTeam system showing: (1) Team Formation phase on the left with team member configuration including four members - Minami as team lead connected to Robin, Charlie, and Diana, displaying member attributes like name, age, gender, major, and skills; (2) Team-Based Ideation phase in the center showing real-time chat between team members discussing and evaluating a 'Smart Vertical Garden Tower' idea for urban gardening systems; (3) Team Reflection phase on the right displaying team information, activity timeline, member details, and generated ideas with their scores. The interface demonstrates the cyclical flow from forming teams, conducting collaborative ideation, to reflecting on team performance.]{Overall flow of \sysname{}. \sysname{} supports three repeating phases—forming HMATs, ideating within HMATs, and reflecting on the team's ideation.}
    \label{fig:crafteam_overall_concept}
\end{center}
\end{figure*}

\section{Design of \sysname{}}
Our primary goal is to investigate what practical challenges and design considerations arise when forming HMATs for collaborative creative work, particularly where assumptions and strategies from HATs and human-only teams may no longer hold. Since HMAT formation remains underexplored, we adopt a technology probe method~\cite{hutchinson2003technology}, well-suited to eliciting user-grounded empirical insights to inform the design of new technologies. To this end, we developed \sysname{}, a technology probe that enables users to form and collaborate with their own HMATs in creative workflows, revealing how they specify HMAT formations in practice and what considerations arise in doing so. The following sections describe the concept and interfaces of \sysname{}, and implementation details are provided in Appendix~\ref{sec:system_detail}.

\subsection{Overall Concept}

\sysname{} is a web application that enables users to form their own HMATs and engage in team-based ideation sessions. To make this process accessible even to non-developers, \sysname{} lets users configure only the core dimensions of team formation, while the system automatically constructs complete HMATs based on the users' settings. In particular, users can directly configure five dimensions of team formation—\textit{team size, structure, role allocation, member composition, and shared mental models} (detailed in Section~\ref{sec:team_formation}).

To enable users to experience the practical consequences of their formation choices for team collaboration, we let them engage in a collaborative ideation task where they work directly with their HMATs to produce creative outcomes. We chose ideation as the primary task because it inherently involves discussion, knowledge exchange, and the integration of diverse perspectives, which has been a widely examined task for human–AI collaboration~\cite{shen2025ideationweb, nomura2024towards, ghosh2025yes, he2024ai}. To implement the co-ideation task with AI agents, we adopted the human-AI co-ideation framework proposed by Shen et al.~\cite{shen2025ideationweb}, which enables humans and AI to share an idea space and track the evolution of design ideas through structured idea representations.

We then incorporated a post-ideation reflection phase with an interface for reviewing ideation transcripts and analyzing interaction patterns. Because collaboration performance in HMATs is shaped by multiple dimensions of team formation acting together, users can struggle to see how any formation choice relates to the quality of their collaborative experience during ideation. By introducing this phase, we enable users to trace specific choices to observed effects and further support systematic evaluation of how their team formation choices affect collaboration during ideation.

We implemented these components as iterative cycles (Fig.~\ref{fig:crafteam_overall_concept}): (i) forming their own HMATs, (ii) ideating with their teams, (iii) reflecting on the ideation session, and then reforming their team based on insights gained from the reflection. We adopted this iterative process to uncover which team formation users select or adjust and how those choices affect collaboration, because it is hard to compare how individual dimensions or their combinations influence collaboration from a single trial~\cite{lykourentzou2016team}. In this study, users were asked to iterate through three cycles, during which they refined their HMAT formation as new empirical insights emerged. These iterations allow us to observe how users' mental models evolve and which team formation they prioritize after gaining hands-on experience.

\subsection{HMAT Formation in \sysname{}}

\begin{figure*} []
\begin{center}
    \includegraphics[width=\textwidth]{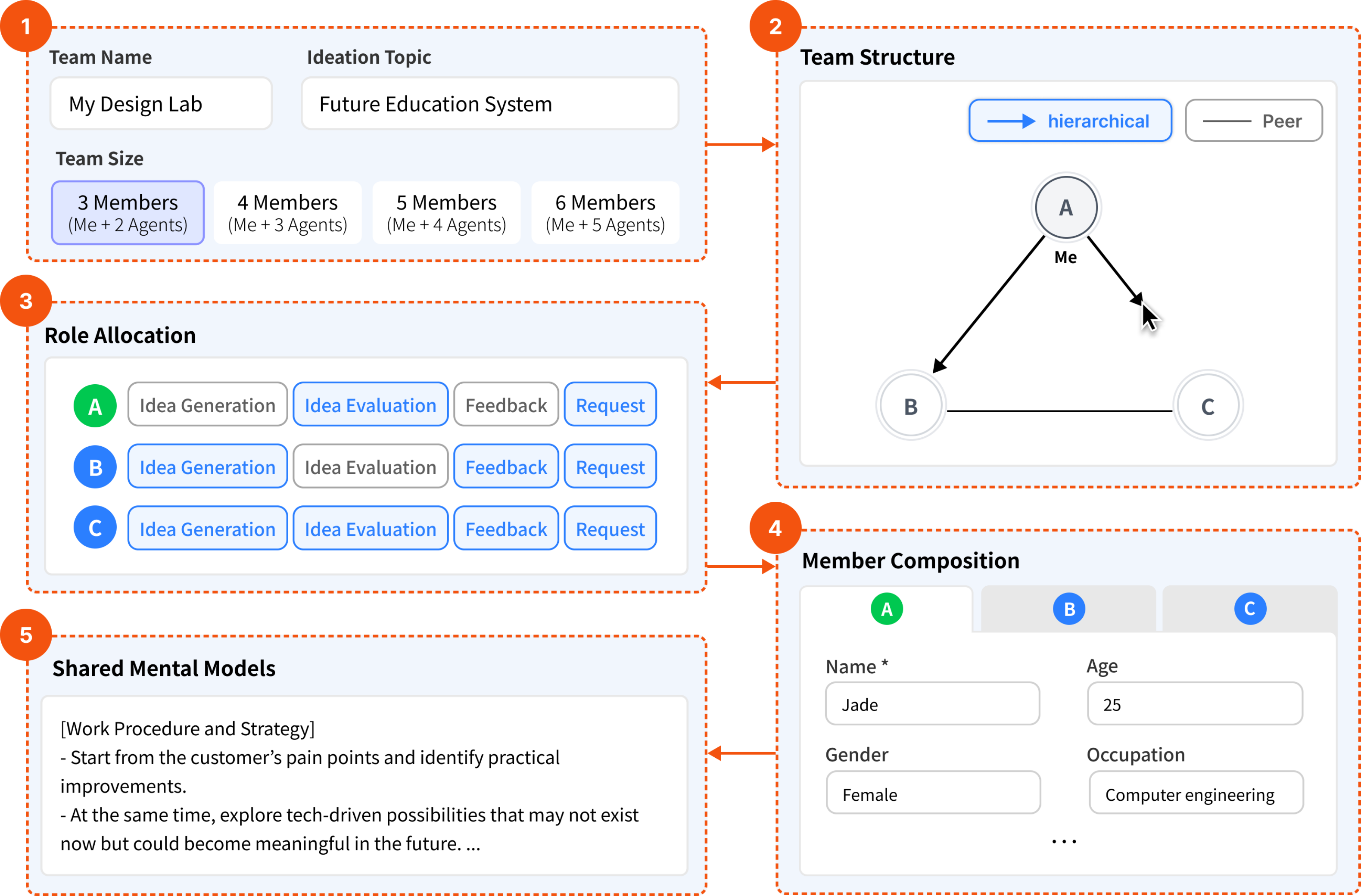}
    \caption[CrafTeam's Team Formation Interface with five numbered configuration panels: (1) Team setup panel showing team name `My Design Lab', ideation topic `Future Education System', and team size options from 3 to 6 members with user-agent composition details; (2) Team Structure panel displaying hierarchical configuration with user at top connected to three agent members A, B, and C; (3) Role Allocation panel showing four available roles (Idea Generation, Idea Evaluation, Feedback, Request) assigned to each team member with selected roles highlighted in blue; (4) Team Composition panel displaying member persona forms with fields for name, age, gender, and occupation for agent members Jade, B, and C; (5) Shared Mental Models panel containing team's work procedure and strategy text describing customer-focused and tech-driven exploration approaches. Orange arrows connect the panels showing the sequential configuration flow.]{Simplified Team Formation Interface of \sysname{}. (1) The user sets the team name and ideation topic and chooses the team size. (2) The user configures a networked team structure by linking members as hierarchical or peer. (3) The user assigns roles to each member—Idea Generation, Idea Evaluation, Feedback, and Request. (4) The user creates member personas in a resume-like form aligned with roles. (5) The user establishes the team's shared mental models.
    }
    \label{fig:crafteam_team_formation_UI}
\end{center}
\end{figure*}

In this section, we introduce five key dimensions of HMAT formation that can be configured in \sysname{}, and present the Team Formation Interface, which enables users to form their own HMATs by manipulating these dimensions.

\subsubsection{Dimensions for HMAT Formation}
\label{sec:team_formation}
Drawing from prior research on human-agent team formation~\cite{abhinav2023survey, iftikhar2024human}, we adopted five dimensions of team formation that shape team dynamics and influence effective collaboration: \textit{team size, structure, role allocation, member composition, and shared mental models}. In selecting these dimensions, we focused on aspects of team formation that non-developer users could explicitly configure through the interface (e.g., size, structure, roles), and excluded dimensions that are more difficult to operationalize as direct settings, such as interaction style or team intervention protocols.

Team size refers to the number of members (including the user) that form a team. Once size is determined, team structure defines how these members organize to break down complex tasks that exceed any individual's capacity~\cite{james2019team}. Within this structure, role allocation assigns specific work functions to each member, both human and automated~\cite{schulte2016design, figueroa2019automatic}. Beyond functional assignments, member composition considers the diversity of member characteristics and how these differences impact team processes and outcomes~\cite{paruchuri2010effect}. Finally, shared mental models (SMMs) refer to the extent to which team members share a common understanding of team tasks, goals, and members' capabilities, which has been linked to improved coordination and team performance~\cite{robert2023role, schelble2022lets, matthias2017framework}. Based on these interconnected dimensions of HMAT formation, we designed the Team Formation Interface, which allows users to configure and refine them based on empirical insights from team-based ideation sessions.

\begin{figure*} [t]
\begin{center}
    \includegraphics[width=\textwidth]{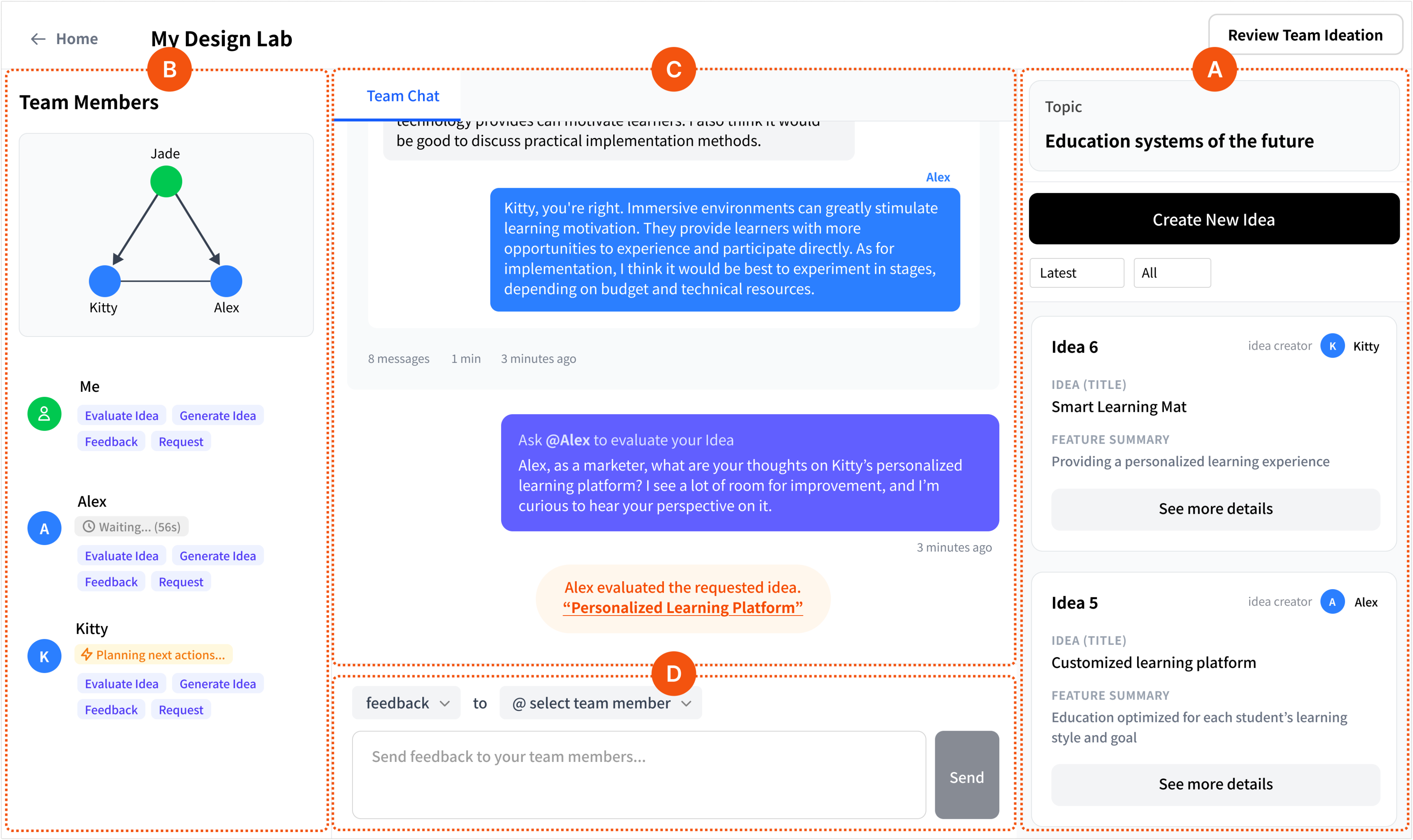}
    \caption[CrafTeam's Ideation Interface with four main components: (A) Idea Tab on the right showing a list of generated ideas including `Smart Learning Mat' by Kitty and `Customized learning platform' by Alex, with `Create New Idea' button and `See more details' options for each idea; (B) Team Members panel on the left displaying hierarchical team structure with Jade at top connected to Kitty and Alex, plus the user's profile showing available actions (Evaluate Idea, Generate Idea, Feedback, Request) and agent status indicators; (C) Team Chat window in the center showing real-time conversation between team members discussing personalized learning platforms and immersive environments, with timestamps and message indicators; (D) Feedback input field at bottom with dropdown menus to select feedback type and target team member, plus Send button. The interface topic shows `Education systems of the future' and enables real-time collaboration between human user and AI agents.]{Ideation Interface of \sysname{}. (A) Idea Tab displays all team-generated ideas. It allows users to generate new ideas through the ``Create New Idea'' button and provides detailed views via the ``See more details'' option, which reveals full idea content and enables updates or evaluations (shown as Fig~\ref{fig:crafteam_idea_representation_UI}). (B) Team Status Tab displays the team structure and member information, including the roles assigned to each member, the relationships between team members, and the current status of the AI agents. (C) Team Chat Window displays real-time interaction logs, including feedback sessions between team members. (D) Chat Input Field enables users to provide feedback or make requests to AI agents.}
    \label{fig:crafteam_ideation_UI}
\end{center}
\end{figure*}

\begin{figure*} []
\begin{center}
    \includegraphics[width=0.7\textwidth]{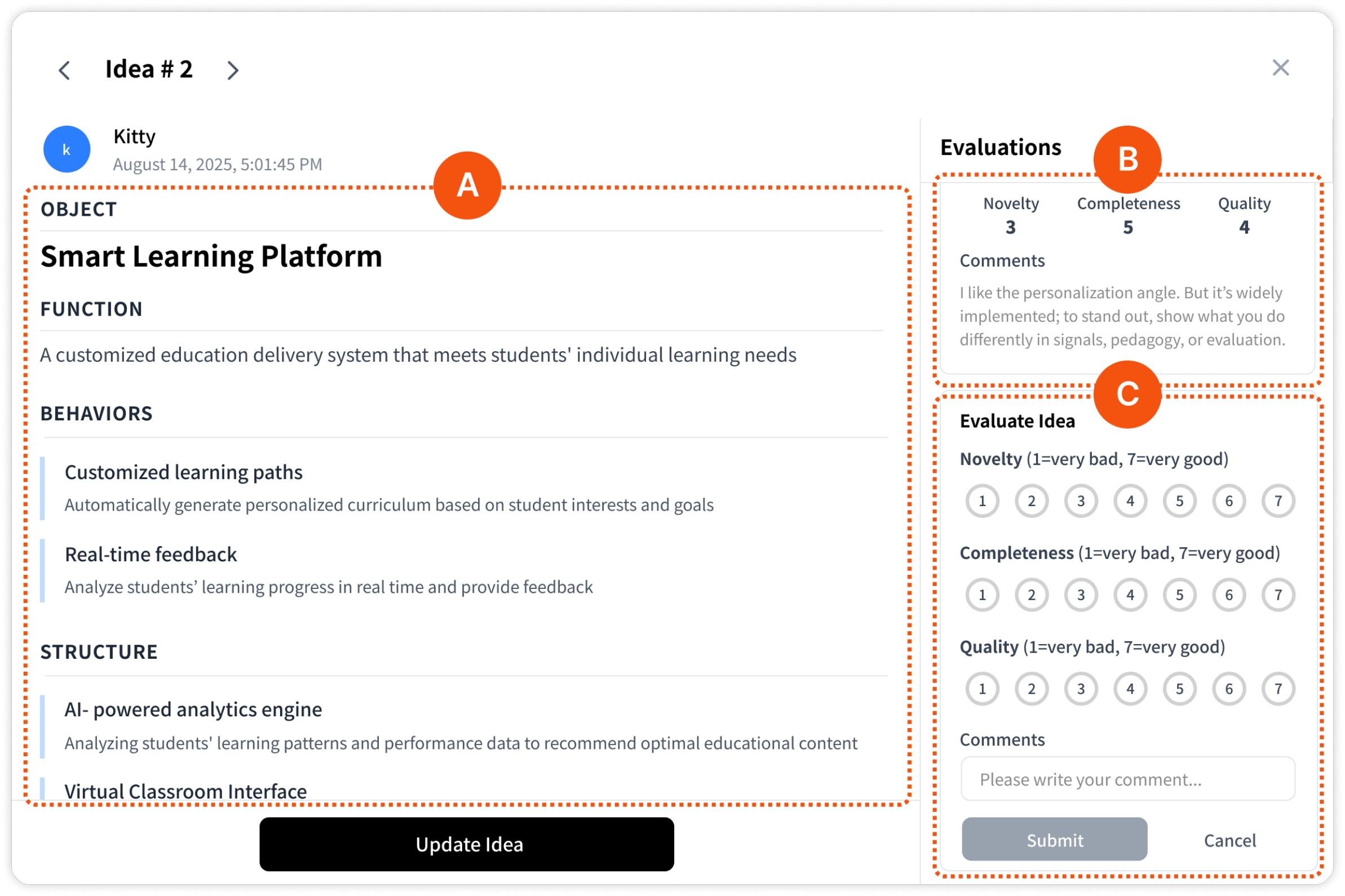}
    \caption[CrafTeam's Idea Card interface displaying three sections: (A) Main idea details panel showing 'Smart Learning Platform' created by Kitty on August 14, 2025, with structured information including Object (Smart Learning Platform), Function (customized education delivery system for individual learning needs), Behaviors (customized learning paths with automated personalized curriculum, real-time feedback analyzing student progress), and Structure (AI-powered analytics engine and Virtual Classroom Interface), with an 'Update Idea' button at bottom; (B) Evaluations section showing existing ratings - Novelty: 3, Completeness: 5, Quality: 4, with a comment stating appreciation for personalization angle but noting need for differentiation; (C) Evaluate Idea form allowing users to rate the idea on three 7-point scales for Novelty, Completeness, and Quality, with a comments field and Submit/Cancel buttons.]{Idea Card in \sysname{}. (A) The main view displays the idea representation, including Object, Function, Behavior, and Structure. (B) The Evaluation List displays team members' ratings and comments on the idea. (C) The Evaluate Idea Tab allows users to provide their own ratings and comments on the idea.}
    \label{fig:crafteam_idea_representation_UI}
    \vspace{-0.2cm}
\end{center}
\end{figure*}

\subsubsection{Team Formation Interface}

The Team Formation Interface is designed to enable users to form their own HMATs by configuring key dimensions of team formation. The interface consists of five stages (Fig.~\ref{fig:crafteam_team_formation_UI}), each corresponding to one core dimension of team formation: 
\begin{enumerate*}[label=(\roman*), itemjoin={{, }}, itemjoin*={{, and }}]
  \item Team Size
  \item Team Structure
  \item Role Allocation
  \item Member Composition
  \item Shared Mental Model
\end{enumerate*}.
After completing all stages, users can finalize their team by clicking the ``Create Team'' button, which then transitions them to the team-based ideation phase.

In the first step, users set basic team information and determine the team size. They begin by naming their team and defining the ideation topic the team will address. Then users set the team size, which can range from 3 to 6 members. We limited the minimum to three members (one user and two AI agents) to ensure multi-agent team interactions could be explored and capped the maximum at six (one user and five AI agents) as an appropriate scale for design ideation and to prevent cognitive overload from managing and interacting with multiple agents~\cite{xu2024adaptation, park2025choicemates}.

In the second step, users establish a team structure by assembling relationships between team members. We implemented a network-based team structure to allow users to create diverse relationships among members and form teams with various structural configurations. Each member serves as a node, and users connect their relationships as edges to create a network graph-style team structure. When establishing relationships, users can choose between hierarchical or peer relationships. Only members who are directly connected can interact with each other.

In the third step, users assign roles to each team member by selecting one or more of four defined roles: Idea Generation, Idea Evaluation, Feedback, and Request (detailed in Section~\ref{sec:action}). Every member must be assigned at least one role, and at least one member must have the Idea Generation role to ensure the team can produce ideas. During the ideation phase, each member can perform only their assigned roles.

In the fourth step, users define the characteristics of each team member by inputting personas in a resume-like format. The information fields are based on the framework for multidimensional identity representation in LLM-based agents~\cite{lee2025spectrum}, including Social Identity (Age, Gender, Education, Occupation), Personal Identity (Personality, Skills), and Personal Life Context (Behavior, Likes, Dislikes), adapted for the context of building ideation teams. We encouraged users to design personas suitable for the intended roles, allowing them to specify only the characteristics they deemed necessary, which enabled us to understand which attributes they prioritized.

In the final step, users establish the team's SMMs. We asked them to write textual descriptions of task models (e.g., procedures, possible outcomes, and how to handle them) and team models (e.g., teammates' tendencies, beliefs, and personalities), which were used as hard-coded guidelines for team-based ideation.

\subsection{Team-Based Ideation of \sysname{}}

After forming HMATs, users move to \sysname{}'s Team-Based Ideation phase to ideate with their team (Fig.~\ref{fig:crafteam_ideation_UI}). We first outline the ideation task and the roles both users and AI agents can take in ideation, then describe the Team-Based Ideation Interface.

\subsubsection{Ideation Task}
In \sysname{}'s team-based ideation phase, users collaborate with their HMATs to engage in a conceptual ideation task on a topic of their choice. In \sysname{}, ideas are represented in a unified format~\cite{shen2025ideationweb} with four components: object (the design target), function (the intended purpose or teleology), behavior (what it does or how it responds, expected or derived from its structure), and structure (its components and their compositional relationships). Each idea is displayed as a card in the interface (Fig.~\ref{fig:crafteam_idea_representation_UI}).

\subsubsection{Roles in Team-based Ideation}
\label{sec:action}
During ideation, the user and AI agents can generate, share, and give feedback on ideas, iteratively refining them through feedback and initial screening. We identified four core roles that both users and agents can take in an ideation session: Idea Generation, Idea Evaluation, Feedback, and Request. Users can serve in these roles through the user interface (Section~\ref{sec:ideation_interface}), whereas AI agents can serve in these roles autonomously via LLM-based pipelines (Section~\ref{sec:agent}). Both users and agents may perform an action only if they were assigned the corresponding role during the team formation phase.

\ipstart{Idea Generation}
Idea Generation is the action of creating idea cards. In this role, members can either (i) \textit{Generate New Idea} to create original concepts or (ii) \textit{Update Idea} to refine existing ideas using their existing representations as templates. All new and updated ideas appear in the Idea tab, which is visible to the whole team (Fig.~\ref{fig:crafteam_ideation_UI}A).

\ipstart{Idea Evaluation}
Idea Evaluation is the action of assessing ideas. In this role, members rate each idea on three 7-point Likert scales —novelty, completeness, and quality~\cite{shen2025ideationweb}—and can optionally add brief comments. Ratings and comments are displayed on the right side of each idea card (Fig.~\ref{fig:crafteam_idea_representation_UI}B).

\ipstart{Feedback}
Feedback is the action of providing conversational feedback to other team members about specific ideas~\cite{lim2025feedometer} or the overall teamwork~\cite{dow2011parallel}. In this role, a member can open a one-on-one chat with a selected team member to whom they are connected in the team structure and exchange multi-turn feedback messages.

\ipstart{Request}
Request is the action of asking other team members to take specific actions in the ideation workflow~\cite{julie1995workload}. In this role, members can request teammates to whom they are connected in the team structure to perform Idea Generation, Idea Evaluation, or Feedback.

\subsubsection{Team-Based Ideation Interface}
\label{sec:ideation_interface}
Figure~\ref{fig:crafteam_ideation_UI} illustrates the Ideation Interface of \sysname{}. The interface supports fluid collaboration by making all team activities visible and accessible, allowing users to respond adaptively to the team's ideation process. 

At the beginning of an ideation session, only Idea Generation is available to establish a shared starting point. After the first idea is generated, they may also perform Idea Evaluation, Feedback, and Request within their assigned role permissions. In the Idea Tab (Fig.~\ref{fig:crafteam_ideation_UI}A), users can click Generate New Idea to add a new idea card. They can also open any idea card in the list to evaluate it or update the idea. Through the Chat Input Field (Fig.~\ref{fig:crafteam_ideation_UI}D), users can send either Feedback or Request by selecting a recipient, choosing the message type (Feedback or Request), and typing a brief message; for Request, they also choose the action for the recipient to perform. Through Team Status (Fig.~\ref{fig:crafteam_ideation_UI}B), users can monitor team- and member-level status to decide when to provide Feedback or make Requests.
Through Team Chat (Fig.~\ref{fig:crafteam_ideation_UI}C), users can see member activity in real time and, once a Feedback conversation ends, review its transcript.

\begin{figure*} [h]
\begin{center}
    \includegraphics[width=\textwidth]{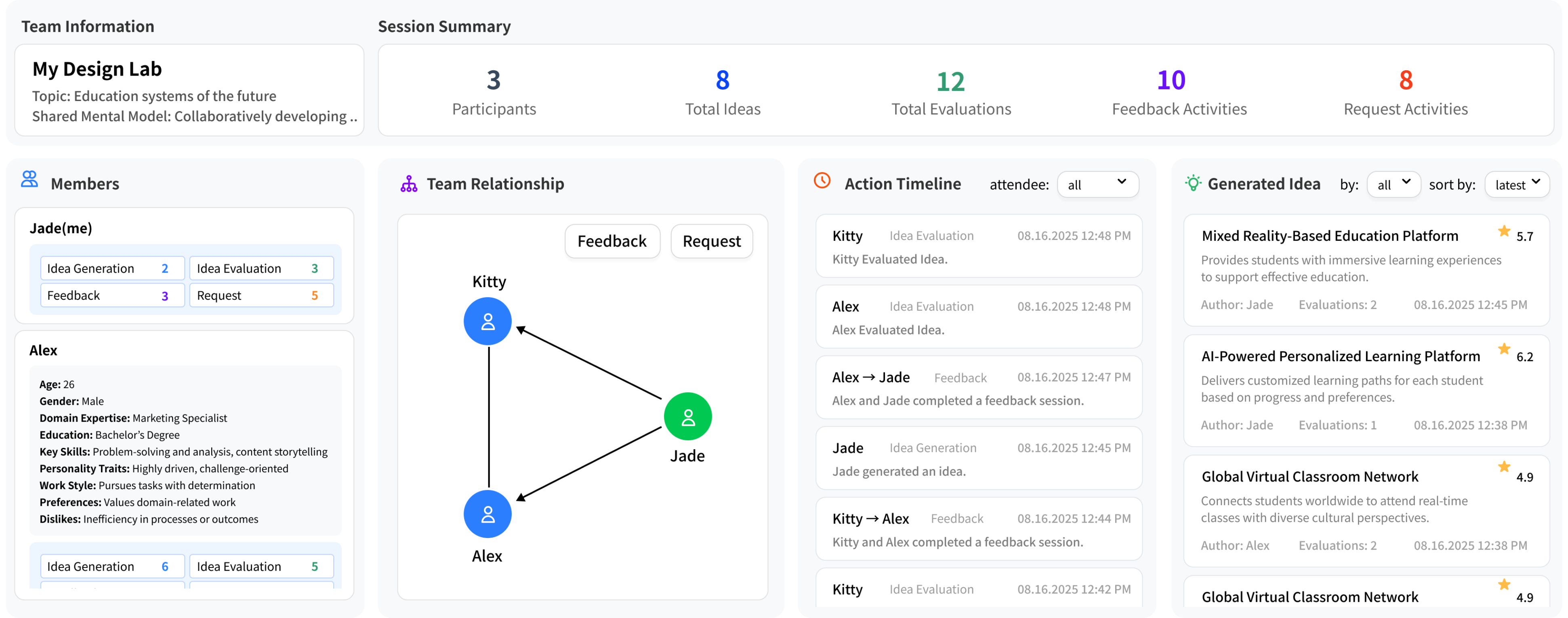}
    \caption[CrafTeam's Reflection Interface displaying five integrated panels: Top section shows Session Summary with team 'My Design Lab' working on 'Education systems of the future' and metrics (3 participants, 8 total ideas, 12 evaluations, 10 feedback activities, 8 request activities). Left side contains Members Panel listing Jade and Alex with their detailed personas and role-specific action counts (e.g., Jade: 2 Idea Generation, 3 Idea Evaluation, 3 Feedback, 5 Request). Center shows Team Relationship Panel with network visualization where Jade connects to Kitty and Alex, with toggle options for Feedback and Request flow visualization. Right side displays Action Timeline Panel showing chronological activities like 'Kitty Evaluated Idea' and 'Alex and Jade completed feedback session' with timestamps. Bottom right shows Generated Ideas Panel listing ideas such as 'Mixed Reality-Based Education Platform' (5.7 rating), 'AI-Powered Personalized Learning Platform' (6.2 rating), and 'Global Virtual Classroom Network' (4.9 rating) with author information and evaluation counts.]{Reflection interface of \sysname{}. The interface consists of five main panels: (i) The Session Summary Panel displays team information and session metrics. (ii) The Members Panel shows each member's persona along with their per-role action counts. (iii) The Team Relationship Panel renders the team network, and when Feedback or Request mode is toggled, it visualizes the flow of these interactions between members. (iv) The Action Timeline Panel presents a chronologically ordered list of actions, featuring an attendee filter and expandable entries for detailed information. (v) The Generated Ideas Panel lists all ideas with their scores and metadata, where each item can be expanded to reveal detailed information.}
    \label{fig:crafteam_reflection_UI}
\end{center}
\end{figure*}

\subsection{Reflection Interface}
\sysname{} provides a reflection phase that enables users to review their ideation sessions with their HMATs and gain insights for team improvement. Once users complete their ideation session, they can proceed to the reflection interface by clicking the ``Review Team Ideation'' button on the Team-Based Ideation Interface. The Reflection Interface (Fig.~\ref{fig:crafteam_reflection_UI}) presents both team-level activity logs and individual member activity logs from the ideation session. Team-level logs show the total number of ideas generated and evaluated, patterns of Feedback and Request exchanges between members, and the final collection of ideas produced. Individual activity logs display how frequently each team member serves within their assigned roles.


%% file: contents/04study.tex
\section{User Study}
In this study, we use \sysname{} to deepen our understanding of HMAT formation by exploring how users form their own HMATs and what considerations and strategies they employ when configuring these teams. Rather than conducting a comparative study to evaluate how well our system supports ideation, we conducted a single-condition study to closely observe creative professionals using \sysname{}, thereby gaining richer insight into how they adapt familiar team formation practices from human-only design teams when forming and working with HMATs for creative workflows. The study protocol was approved by our institution's Institutional Review Board (IRB).

\input{tables/demographic}

\subsection{Participants}
A total of 12 participants (six females, six males) aged from 26 to 35 ($M=29.42, SD=3.82$) were recruited through IT industry communities in South Korea. To ensure participants had sufficient background to form design teams informed by real-world experience, we recruited design practitioners currently engaged in team-based design work at IT companies and with prior experience using AI agents in their workflows. Participants' demographic details are presented in Table~\ref{tab:demographic}. The study was conducted in a 200-minute session, and participants received 200,000 KRW (approximately USD 145) as compensation.

\begin{figure*} []
\begin{center}
    \includegraphics[width=\textwidth]{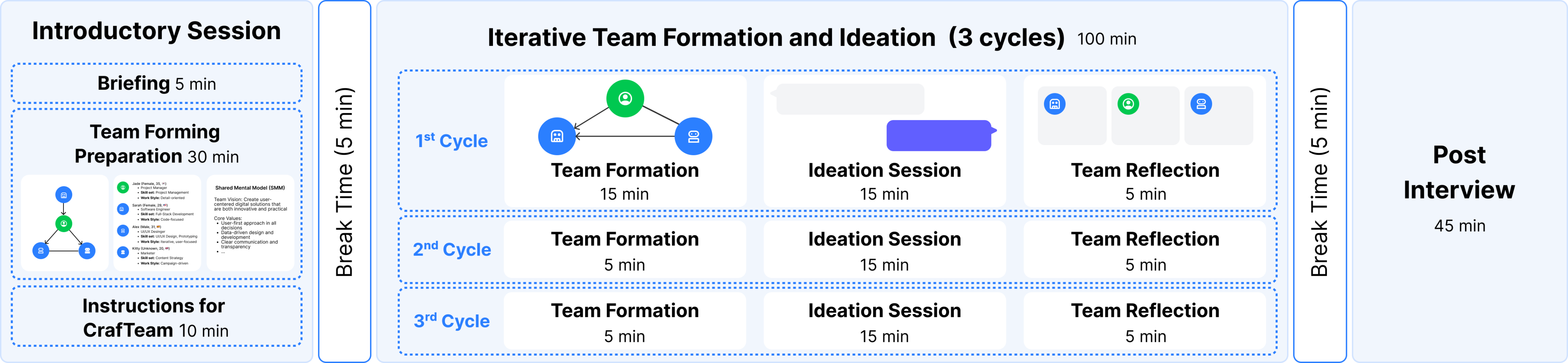}
    \caption[User study process diagram showing three main phases across a timeline. Left: 45-minute Introductory Session consisting of 5-minute briefing, 30-minute team building preparation with a sample team structure diagram and shared mental model document, and 10-minute CrafTeam instructions. Center: 100-minute Iterative Team Building and Ideation section with three cycles, each containing Team Building (15 min in Cycle 1, 5 min in Cycles 2-3), Ideation Session (15 minutes), and Team Review (5 minutes), illustrated with team network diagrams and chat interface icons. Right: 45-minute Post Interview phase. Timeline indicators show 5-minute breaks between major sections. Total study duration approximately 3.5 hours.]{User study process. The study included a 45-minute introductory session, followed by three iterative cycles of team formation, ideation, and reflection (25 minutes each). After the cycles, participants engaged in a 45-minute post-interview to reflect on their experiences and share insights.}
    \label{fig:study_protocol}
\end{center}
\end{figure*}

\subsection{Procedure}
The user study process is illustrated in Figure~\ref{fig:study_protocol}. Each study session lasted approximately three hours and consisted of four phases: (i) Introductory session, (ii) Iterative team formation and ideation, and (iii) Post-interview.

\ipstart{Introductory session}
We first explained the purpose and procedure of the study to participants and obtained their signed consent. We informed participants that they would be composing teams with AI agents to perform ideation tasks and provided a brief tutorial on system usage.

Before using the \sysname{}, we conducted a team forming preparation session to help participants engage with HMAT formation. Participants first answered questions about their past creative teamwork experiences (e.g., most effective team experiences, their roles in teams) and their conception of ideal team conditions. They then completed three worksheets: (i) sketching an ideal team structure, (ii) creating profiles for desired team members, and (iii) defining the team's shared mental models. These activities helped participants reflect on HMATs' team formation and develop concrete ideas about the teams they wanted to implement in \sysname{}.

\ipstart{Iterative Team Formation and Ideation}
After the introductory session, we introduced participants to the \sysname{} and its interfaces. We then asked them to complete three cycles of team formation, ideation, and reflection using \sysname{}. In the first team formation phase, we instructed participants to form a new HMAT based on their preparatory worksheets and freely choose topics for ideation from their professional experiences (15 minutes). In the second and third cycles of team formation, participants reconfigured their teams by modifying their previous setups (5 minutes each). For each ideation phase, participants conducted ideation with their newly formed team for 15 minutes. Before beginning the ideation, we emphasized that the primary goal was not to generate perfect ideas in the short time available, but to use the session as an evaluation to consider how to form an effective ideation team. Finally, after each ideation, we guided participants through a reflection phase, prompting them to evaluate team effectiveness and identify areas for improvement using system-provided activity logs.

\ipstart{Post-Interview}
We concluded with 45-minute semi-structured interviews to explore participants' experiences in depth. The interview covered: insights gained from iteratively developing teams across three cycles, changes in perception regarding the importance of each dimension of team formation, experiences collaborating with AI agents, and suggestions for future HMAT design. All interviews were audio-recorded for transcription and analysis.

\subsection{Data Analysis}
We first conducted a descriptive statistical analysis to examine how participants formed and revised their teams. To analyze participants' trial-and-error processes, we quantified changes across cycles for the five dimensions of team formation that participants could adjust. To better understand these changes in team structure, we categorized the teams through iterative comparison, resulting in three distinct types (Fig.~\ref{fig:team_structure}).

To gain a deeper qualitative understanding, we conducted open coding and thematic analysis~\cite{virginia2006thematic}. In particular, our analysis aimed not to highlight considerations known from forming one-human-one-agent teams, but to identify unique considerations for forming HMATs, as well as the requirements of participants. The first PhD author open-coded the interview transcripts and interaction log data through multiple rounds of iteration. Two additional PhD researchers then reviewed the initial themes and supporting quotes, providing feedback on the coding. Based on this, the entire research team iteratively revised the themes, ultimately producing enhanced qualitative findings that captured both the factors participants considered when forming HMATs and the requirements they identified from their perspective.

%% file: tables/demographic.tex
\begin{table}[h]
\sffamily
    \renewcommand{\arraystretch}{1.4}
    {\footnotesize
    \begin{tabular}{ccccc}
        \toprule
        \textbf{\begin{tabular}[c]{@{}c@{}}Participant \\ ID\end{tabular}} & \textbf{Age} & \textbf{Gender} & \textbf{Occupation} & \textbf{Domain} \\ \toprule
        P1 & 35 & M & Service Designer & AI startup\\ \hline
        P2 & 26 & F & UI/UX Designer & Screen Interface\\ \hline
        P3 & 27 & M & UX Designer & Searching Platform\\ \hline
        P4 & 26 & F & UX Designer & Advertisement\\ \hline
        P5 & 28 & M & Service Designer & Searching Platform\\ \hline
        P6 & 26 & F & UX Designer & VR/AR\\ \hline
        P7 & 32 & M & Project Manager & Video Game\\ \hline
        P8 & 35 & F & UX Designer & Video Game\\ \hline
        P9 & 26 & F & UI/UX Designer & Screen Interface\\ \hline
        P10 & 33 & M & UX Designer & AI startup\\ \hline
        P11 & 33 & M & Product Manager & Matchmaking Platform\\ \hline
        P12 & 26 & F & UX Designer & VR/AR\\ \bottomrule
    \end{tabular}
    }
    \vspace{0.2cm}
    \caption[Demographic table showing 12 study participants (P1-P12) with columns for Participant ID, Age (ranging from 26-35), Gender (6 male, 6 female), Occupation (including 7 UX Designers, 2 UI/UX Designers, 2 Service Designers, 1 Project Manager, and 1 Product Manager), and Domain expertise (including AI startup, Screen Interface, Searching Platform, Advertisement, VR/AR, Video Game, and Matchmaking Platform). The majority of participants were UX/UI designers in their late twenties to mid-thirties working across various digital product domains.]{Demographic information of study participants.}
    \label{tab:demographic}
    \vspace{-0.5cm}
\end{table}

%% file: contents/05finding.tex
\section{Findings}

\input{tables/team_formation}

In this section, we first present descriptive statistics on how participants form HMATs and conduct ideation with their teams using \sysname{}. We then describe the key considerations participants encountered when forming teams and the strategies they used to address them. Finally, we report the requirements for human-multi-agent teaming that participants identified, distinguishing it from both human-only and one-human-one-agent teams.

\subsection{Descriptive Statistics of \sysname{} Usage}
We first describe how participants formed their teams, focusing on the HMAT formation dimensions they specified, and then report how ideation unfolded within these teams. To shed light on the trial-and-error process through which participants formed their teams, we report how teams evolved across cycles. We denote participants as P1, and index each team by the cycle number following the participant ID (T1–T3). For example, P1T3 refers to P1's team in Cycle 3.

\subsubsection{Participants' HMAT Formation Patterns}
In total, participants formed 36 teams (12×3 cycles), exploring diverse formations—parti-cularly in team structure and role allocation. Table~\ref{tab:team-formation-cycles} summarizes how participants configured the dimensions of team formation across cycles.

\ipstart{Team Size}
Participants created teams averaging 4.61 members ($SD = 1.02, min = 3, max = 6$), including themselves. They initially formed teams averaging 5.00 members, scaled them down to 4.08 in the second cycle, and then rebounded to 4.75 in the final cycle.

\begin{figure*} []
\begin{center}
    \includegraphics[width=\textwidth]{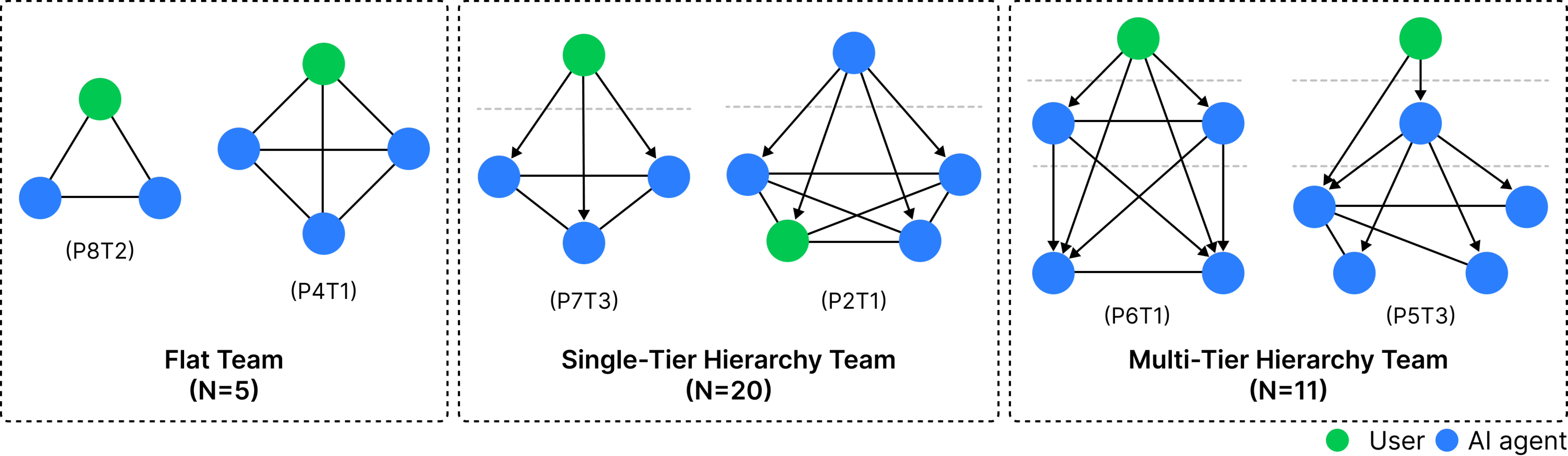}
    \caption[Diagram showing three categories of team structures from 36 participant-built teams. Left panel: Flat Teams (N=5) illustrated by two network diagrams (P8T2 and P4T1) where green user nodes connect to blue AI agent nodes in non-hierarchical arrangements with horizontal connections between agents. Center panel: Single-Tier Hierarchy Teams (N=20) shown through two examples (P7T3 and P2T1) where a single green user node at the top directly manages all blue AI agents below, with varying degrees of inter-agent connections. Right panel: Multi-Tier Hierarchy Teams (N=11) demonstrated by two structures (P6T1 and P5T3) featuring green user nodes at the top with blue AI agents arranged in multiple management layers, showing cascading hierarchical relationships. Legend indicates green circles represent users and blue circles represent AI agents.]{Three types of team structures emerged from teams formed by participants: (i) Flat teams where all members functioned at the same hierarchical level; (ii) Single-tier Hierarchy where one or a few leaders directly managed all other members; and (iii) Multi-tier Hierarchy where multiple layers of management existed between top leadership and base members.}
    \label{fig:team_structure}
\end{center}
\end{figure*}

\ipstart{Team Structure}
We categorized team structures into three types based on how participants configured hierarchical relationships (Fig.~\ref{fig:team_structure}): (i) Flat, (ii) Single-tier Hierarchy, and (iii) Multi-tier Hierarchy.

Single-tier Hierarchy teams were most common, with the user typically serving as the sole leader managing all agents directly. Only P2T1–T3 placed an AI at the top, and only P1T2 and P1T3 assigned multiple co-leaders. Multi-tier teams comprised a top manager, one or two mid-level managers, and several team members; users generally occupied the top role while AI agents served as mid-level managers, with the sole exception of P3T1, where the participant took a mid-level position. By the end of the study, participants tended to converge on Single-tier Hierarchy structures with themselves as the leader. This pattern indicates an initial exploration of various structural options, followed by a pragmatic preference for direct control, in which the user maintains a clear leadership role over all agents.

\ipstart{Role Allocation}
Participants assigned averaged 2.99 roles per member. For AI agents, the number assigned to each role per team decreased from Cycle 1 to 2, then partially recovered in Cycle 3. Teams assigned an average of 3.58 AI agents to Idea Generation in Cycle 1, reduced to 2.42 in Cycle 2, then increased to 3.08 in Cycle 3. Idea Evaluation followed the same pattern with averages of 3.50, 1.92, and 2.58 agents. Feedback roles decreased from an average of 3.83 agents to 2.00, then partially recovered to 2.75. Request roles remained stable.

Users rarely participated in Idea Generation (36\% of teams) but frequently took Idea Evaluation roles (81\% of teams). All teams included users in the Feedback and Request roles.

\ipstart{Member Composition}
Participants created 130 AI agent profiles to compose their HMATs. Completion rates across the three categories were: Personal Identity 96.96\% (skills 100\%, personality 94\%); Social Identity 90.96\% (age 90\%, gender 87\%, education 87\%, occupation 100\%); and Personal Life Context 83.08\% (work style 94\%, likes 78\%, dislikes 78\%). In composing teams, participants prioritized occupations and skill sets, while attributes such as gender and likes/dislikes were comparatively non-essential.

\ipstart{Shared Mental Model}
Participants established SMMs averaging 203.83 syllables in length. Participants entered longer SMMs in Cycle 1 (226.58 syllables), then progressively shorter SMMs in Cycle 2 (194.67 syllables) and Cycle 3 (190.25 syllables), demonstrating a decreasing trend across cycles.

\subsubsection{Patterns of Actions in Team-Based Ideation}
Across three team-based ideation sessions, participants engaged in ideation with their HMATs on a variety of topics, including future wearable AI devices (P1), a context-aware Next TV platform UI (P6), and conversational AI robots for children's emotional development (P9). In this section, we highlight key patterns in how participants and AI agents differed in their actions across the three ideation cycles, and present a detailed summary of these patterns in Appendix~\ref{sec:ideation_detail}.

\ipstart{Idea Generation}
HMATs generated 451 ideas in total. Most of these ideas came from AI agents, whereas participants rarely generated ideas directly. Across teams, AI agents with the Idea Generation role produced an average of 4.37 ideas, whereas participants produced only 0.19 ideas (7 ideas in total). The volume of ideas remained similar across cycles (151, 146, and 154 ideas in Cycles~1–3), but each member's idea productivity shifted over time. The number of ideas generated by each member increased from an average of 3.49 in Cycle 1 to 5.07 in Cycle 2, when generation roles were concentrated on fewer agents, and then decreased slightly to 4.54 in Cycle 3 after roles were rebalanced.

\ipstart{Idea Evaluation}
HMATs produced 428 evaluations total. Overall, participants gave relatively low but gradually increasing ratings across cycles, with substantial variability across ideas, from $M = 3.12 \,(SD = 1.66)$ in Cycle 1 to $M = 4.06 \,(SD = 1.66)$ in Cycle 2 and $M = 4.51 \,(SD = 1.37)$ in Cycle 3. In contrast, AI agents consistently gave high and similar ratings across cycles, scoring $M = 5.36 \,(SD = 0.61)$ in Cycle 1, $M = 5.37 \,(SD = 0.50)$ in Cycle 2, and $M = 5.37 \,(SD = 0.43)$ in Cycle 3, indicating that most ideas were evaluated within a narrow high range.

\ipstart{Feedback}
HMATs conducted 358 feedback sessions, averaging 5.62 turns each. AI agents initiated 113 sessions in Cycle 1, dropping to 50 in Cycle 2, before partially recovering to 84 in Cycle 3. In contrast, participants became more active in giving feedback across cycles, from an average of 1.50 sessions in Cycle~1 to 2.50 in Cycle~2 and 3.17 in Cycle~3.

\ipstart{Requests}
HMATs made 106 requests total. Participants overwhelmingly requested Idea Generation (49 requests), while only 15 requested Idea Evaluation and 8 requested Feedback. By contrast, AI agents showed a different distribution of requests: only 1 for Idea Generation, 12 for Idea Evaluation, and 21 for Feedback.

\subsection{Participants' Considerations for HMATs Formation}
During the preparation session, participants defined what they wanted their HMATs to achieve and how they expected them to operate. Most sought to enable high-quality ideation while minimizing their own workload, expecting agents to autonomously generate and elaborate ideas with minimal human intervention. During ideation, however, participants found that their teams did not operate as expected and incrementally refined their teams through iterative cycles. In this section, we illustrate the key considerations participants had to address when forming HMATs that differed from their initial expectations, and how they responded to these challenges.

\subsubsection{Delegating Idea Generation to AI Agents}
During team formation, all participants assigned the Idea Generation role to AI rather than generating ideas themselves. They viewed AI as more efficient because brainstorming and writing descriptions are cognitively taxing and time-consuming for humans, whereas AI can produce ideas instantly. They also highlighted reduced social pressure in team ideation, noting that AI produces candidate ideas immediately without the self-censorship that slows humans. As P12 noted, \textit{``In real-world team ideation, people take a long time to voice ideas because they hesitate or resist having them evaluated by teammates. AI, however, provides ideas immediately without that burden, making it well-suited to the ideation role.''}

Most participants believed that delegating ideation to AI agents was efficient in terms of work productivity; some participants (P5, P8, P10, and P11) questioned whether this division of roles actually yields high-quality ideas. They noted that many AI-generated ideas lacked the novelty or practical feasibility needed for real-world application: \textit{``I often use GPT for ideation and have a sense of what generative AI can do; what I see here is similar. For now, humans are better suited for creativity and should handle the truly creative and productive aspects. (P11)''} 

\subsubsection{AI Leaders Fall Short: Users Forced into Direct Management}
Participants said that assigning a leader is necessary to keep multi-member teams aligned toward a unified direction. In this view, they considered themselves better suited to assume this role in HMATs, while also expecting that if AI could replace or support leadership, teams could operate more autonomously with minimal human intervention. Many participants attempted to deploy higher-level or middle-manager agents to assist with, or even fully automate, such management tasks.

However, participants reported that the AI leader did not fulfill the expected leadership role and did not improve the quality of ideas. They mainly attributed this to the agent's reluctance to take clear positions or make decisions. As P5 stated, \textit{``In practice, meetings with a leader end with action items about what to do next and how to proceed. But the AI gave feedback without actually deciding anything and showed no clear preferences; it just kept the conversation going. As a result, ideas did not develop in a specific direction, and we felt like we were going in circles.''} Participants also noted that management functions—such as task allocation and support for individual members—were not performed: \textit{``I expected that when I made a request to the AI set as the leader, it would notify other team members as appropriate and assign tasks suited to each person's abilities. It did not. (P12)''}

Consequently, most participants gradually adopted a single-tier hierarchy, with themselves serving as the leader and directing the team. However, managing multiple agents alone proved burdensome, adding substantial workload: \textit{``Even with only three AIs, whenever a feedback request came in, I had to double-check their prior work; when another agent requested feedback, I had to review that history too. As a result, keeping up was difficult because the agents moved quickly, while my capacity to interpret and provide feedback simultaneously was limited. (P3)''} Ultimately, participants either limited the number of agents or devised more effective management strategies. 

\subsubsection{Generalist to Specialist}
Participants noted that teams where everyone's contributions were respected tended to achieve better results. In line with this view, they designed HMATs to support recognizing contributions and ensuring autonomy, enabling all members to generate ideas, evaluate them, and provide feedback rather than enforcing rigid, role-based divisions dictated by the AI structure.

However, after ideating with such AI teams, participants found that giving a single agent multiple roles slowed task progress and was inefficient. P9 remarked, \textit{``When I initially allowed agents to handle every role, they performed each only superficially. As I clarified roles, for example, ideation versus feedback, the outcomes became more numerous and specific, so I assigned some agents to ideation only and others to feedback only.''} Furthermore, because the structure placed the user in the manager role overseeing every AI agent, assigning multiple roles to each agent increased the coordination workload of participants. P12 explained, \textit{``Initially, I gave them a high degree of freedom, but it was hard to track what each agent had done. So, rather than multiple roles, assigning each agent a single role proved more effective for keeping agents on track.''}

Through iterative cycles, participants converged on assigning each agent a clearly defined role, most often dividing the team into ideation and evaluation. In turn, they emphasized the ability to operate with more extreme specialization, a possibility unique to AI teams, beyond what small, human-only teams can manage. P4 noted, \textit{``In human teams, assigning roles like `you can't ideate, so just evaluate' raises equity concerns, which makes this hard to do. With AI, those constraints don't apply, so if efficiency is the goal, even extreme role separation seems possible.''} Several participants (P1, P5, P7) likened HMAT configuration to a strategy game and approached assignments from a team-first perspective rather than tailoring them to individual needs. P1 further argued for maximizing AI speed and uptime by predefining algorithms, operating policies, and action-selection rules to reduce real-time decision overhead.

\subsubsection{More Agents do not guarantee Better Ideas: Diversifying Agents' Persona}
Participants expected that adding more agents would increase productivity when forming HMATs. Most assigned two or more agents to Idea Generation and Evaluation. However, they found that output did not scale linearly, since agents often produced overlapping ideas or similar evaluations. The agents seemed to lack distinct perspectives, operating more like duplicates than diverse team members: \textit{``At first, it felt less like several people brainstorming together and more like running three GPTs in parallel and assigning tasks to each.  The only benefit seemed to be faster output through parallel processing, not better ideas. (P8)''}

During ideation, participants found that agents with different personas tended to generate ideas aligned with their backgrounds. In response, they diversified the personas of agents in the same role to broaden the scope of ideas. For example, P8 said, \textit{``The agent with a designer persona seemed to focus on more design-oriented ideas, while the developer agent appeared to strive for more technical solutions. So in the next cycle, I added an agent with a civil servant persona and asked it to generate ideas related to national policy.''} Some participants also created personas that do not exist in reality to elicit more unusual ideas. P7 suggested,  \textit{``To add variation to the team, it might be good to intentionally include an extreme member, something like a virus. I instilled a mindset such as `You only think about this direction and are only interested in these things,' and when it interacts with other agents, it might lead to more unexpected outcomes.''} Similarly, for idea evaluation, some participants (P3, P5, and P11) differentiated evaluator agents into personas that provided only positive feedback and those that provided only negative feedback, which yielded evaluations from more diverse perspectives.

\subsection{User-Centered Requirements for Human–Multi-Agent Teaming}
Through \sysname{}, participants had the opportunity to form teams of multiple AI agents and explore how this affected their ideation. All participants indicated that HMATs have potential in their practice and could be applied to other tasks beyond ideation. However, they also identified requirements distinct from human-only teams, underscoring what must be addressed for effective collaboration with multiple agents. In this section, we outline the requirements participants identified while using \sysname{}.

\subsubsection{Challenges Around Interaction with Multi-Agents}
\strut\\
\ipstart{Inefficiency of Human-Like Communication Between Agents}
In \sysname{}, participants could monitor how agents communicated with one another and how well each agent performed its assigned role. With this visibility, they could identify moments when agents behaved differently from their expectations and make fine adjustments. For example, P4 noted, \textit{``I observed the agents exchanging feedback, but they were focusing too much on security issues. Since it was still the early stage of ideation, I told them to focus more on novelty in their feedback, and afterwards, they seemed to provide feedback more in line with what I had expected.''}

However, several participants (P1, P5, and P11) questioned the premise that agent-to-agent communication should mimic human conversation. P11 said, \textit{``If AI agents are conversing among themselves, the exchange need not be in natural language, nor always be visible to me. As it is, they seem to mimic humans, which creates a slight uncanny valley effect.''} P1 argued that human-style dialogue can waste time and computational resources: \textit{``Generative AI typically produces content faster than people, but conversation is different. Because each turn waits for the other's output, it can feel slower than human conversation, and forcing AI to converse this way may be inefficient.''} P5 emphasized the need to revise how AI-to-AI dialogue is surfaced to users: \textit{``I read some AI-to-AI conversations, but they were too noisy, and going through them felt like a waste of time. It would be better to show a summary or simply the outcome: the conclusion they reached.''}

\ipstart{Needs for Team-level Interaction}
Many participants (P2, P5, P6, P8, P10, and P12) mentioned a need for team-level interactions during group ideation that \sysname{} did not support. As P1 noted, \textit{``In real work, we don't just stack one-on-one chats; we need group discussions where several people can talk at once. I wish the system supported that.''} Some participants repurposed the shared mental model as a broadcast channel, adding guidance they wanted to disseminate, such as `Don't suggest ideas for specific technologies' or `Focus on IoT-based services.' They also wanted to integrate these broadcast instructions mid-session so that all agents would adjust in real time during the ideation. Participants further requested a more dynamic, multi-party exchange with overlapping contributions and rapid floor shifts, rather than strictly sequential turns. P6 said, \textit{``When brainstorming with people, everyone takes turns, but talk still overlaps. When two are speaking, others aren't only observing. Three may jump in, then four, then it returns to two. I wish those transitions felt more natural.''} They especially wanted to join ongoing agent-to-agent exchanges at any moment, including interrupting or steering them.



\subsubsection{Rethinking Team Growth in HMATs}
Drawing on real-world workplace experience, participants treated a team's potential for sustained growth as one of the key considerations in team formation. The most common strategy was adapting master–apprentice pairings to HMATs. P6 explained, \textit{``Senior–junior pairs work well in practice because seniors benefit from assistance, and juniors can develop over time through that collaboration.''} P10 also experimented with a competitive structure to develop and refine ideas, dividing the team into two competing groups with the expectation that both would improve over time.

Participants viewed these strategies as less effective for AI agents, arguing that HMATs require different team-development approaches than human-only teams. P8 noted, \textit{``With human teams, once people are hired, it's difficult to dismiss them, so it's crucial to help selected members grow. With AI, you can simply swap in a better model, so investing in a weak agent is less necessary.''} Consequently, participants suggested focusing less on cultivating specific agents and more on iteratively replacing them to better fit the team's structure and role requirements. For example, P1 proposed: \textit{``If we simulate 100 AI teams, award points to teams that generate strong ideas, and then select the top performer, akin to reinforcement learning, we could identify an optimal AI team.''}

%% file: tables/team_formation.tex
\begin{table*}[t]
\sffamily
\centering
\small
\setlength{\tabcolsep}{4pt}
\begin{tabular}{l l r r r r}
\toprule
\textbf{Dimension} & \textbf{Metric} & \textbf{Cycle 1} & \textbf{Cycle 2} & \textbf{Cycle 3} & \textbf{Total} \\
\toprule
\multirow{1}{*}{Team Size}
 & Size (M$\pm$SD)                                 & 5.00$\pm$0.82 & 4.08$\pm$0.95\,\dec & 4.75$\pm$1.01\,\inc & 4.61$\pm$1.02 \\
\midrule
\multirow{3}{*}{Team Structure}
 & Flat Team (N)                                   & 2 & 2\,\nc & 1\,\dec & 5 \\
 & Single-tier Hierarchy Team (N)                  & 6 & 6\,\nc & 8\,\inc & 20 \\
 & Multi-tier Hierarchy Team (N)                   & 4 & 4\,\nc & 3\,\dec & 11 \\
\midrule
\multirow{12}{*}{Role Allocation}
 & All: roles per member (M$\pm$SD)       & 3.22$\pm$0.75 & 2.82$\pm$1.02\,\dec & 2.91$\pm$1.06\,\inc & 2.99$\pm$0.97 \\
 \cmidrule(lr){2-6}
 & Agents: roles per member (M$\pm$SD)             & 3.27$\pm$0.76 & 2.68$\pm$1.07\,\dec & 2.82$\pm$1.12\,\inc & 2.95$\pm$1.02 \\
 & Agents in role: Idea Generation (M$\pm$SD)      & 3.58$\pm$0.95 & 2.42$\pm$0.64\,\dec & 3.08$\pm$0.95\,\inc & 3.03$\pm$0.99 \\
 & Agents in role: Idea Evaluation (M$\pm$SD)      & 3.50$\pm$1.44 & 1.92$\pm$1.04\,\dec & 2.58$\pm$1.04\,\inc & 2.67$\pm$1.35 \\
 & Agents in role: Feedback (M$\pm$SD)             & 3.83$\pm$2.00 & 2.00$\pm$1.29\,\dec & 2.75$\pm$1.64\,\inc & 2.86$\pm$1.57 \\
 & Agents in role: Requests (M$\pm$SD)             & 2.17$\pm$1.67 & 1.92$\pm$1.11\,\dec & 2.17$\pm$1.34\,\inc & 2.08$\pm$1.40 \\
 \cmidrule(lr){2-6}
 & User: roles per member (M$\pm$SD)               & 3.00$\pm$0.71 & 3.25$\pm$0.72\,\inc & 3.25$\pm$0.72\,\nc & 3.17$\pm$0.73 \\
 & User in role: Idea Generation (N)               & 3 & 5\,\inc & 5\,\nc & 13 \\
 & User in role: Idea Evaluation (N)               & 9 & 10\,\inc & 10\,\nc & 29 \\
 & User in role: Feedback (N)                      & 12 & 12\,\nc & 12\,\nc & 36 \\
 & User in role: Requests (N)                      & 12 & 12\,\nc & 12\,\nc & 36 \\
\midrule
\multirow{5}{*}{Member Composition}
 & User (\% attributes specified) & 87.96\% & 87.96\%\textemdash & 87.96\%\textemdash & 87.96\% \\
 \cmidrule(lr){2-6}
 & Agents: Social Identity (\% attributes specified) & 90.10\% & 89.19\%\,\dec & 93.33\%\,\inc  & 90.96\% \\
 & Agents:Personal Identity (\% attributes specified) &  96.88\% & 97.30\%\,\inc & 96.67\%\,\dec & 96.92\% \\
 & Agents:Personal Life Context (\% attributes specified) & 82.64\% & 83.78\%\,\inc & 82.96\%\,\dec & 83.08\% \\
\midrule
\multirow{1}{*}{Shared Mental Model}
 & Text length (syll.; M$\pm$SD)                   & 226.58$\pm$108.47 & 194.67$\pm$100.81\,\dec & 190.25$\pm$109.23\,\dec & 203.83$\pm$107.47 \\
\bottomrule
\end{tabular}
\vspace{0.2cm}
\caption[Comprehensive table showing team dynamics statistics across three cycles with five main dimensions. Team Size averaged 5.00±0.82 in Cycle 1, decreased to 4.08±0.95 in Cycle 2, then increased to 4.75±1.01 in Cycle 3. Team Structure distribution showed 5 Flat Teams, 20 Single-tier Hierarchy Teams, and 11 Multi-tier Hierarchy Teams total. Role Allocation section details roles per member for all participants (2.99±0.97 average), with separate breakdowns for AI agents and users across four roles (Idea Generation, Evaluation, Feedback, Requests). Team Composition shows attribute completion rates: Users maintained 87.96\% across all cycles, while Agents varied - Social Identity 90.96\%, Personal Identity 96.92\%, Personal Life Context 83.08\% overall. Shared Mental Model text averaged 203.83±107.47 syllables, decreasing from 226.58 in Cycle 1 to 190.25 in Cycle 3. Table includes directional markers indicating increases (▲), decreases (▼), and no change (—) between cycles.]{Descriptive statistics of participants' team formation dimensions across three cycles. Markers show change vs.\ previous cycle: \protect\inc{} increase, \protect\dec{} decrease, \protect\nc{} no change.}
\label{tab:team-formation-cycles}
\vspace{-0.5cm}
\end{table*}

%% file: contents/06discussion.tex
\section{Discussion}
In this study, we investigated how participants formed HMATs and collaborated on creative ideation tasks. Through \sysname{}, participants initially attempted to form autonomously operated teams in which AI agents collectively assumed both generative and reflective roles. However, our findings reveal that, because AI agents who have to lead the team struggled to provide value judgments and set directions, which are essential for idea development, participants shifted to team formations in which they directly orchestrated the agents and guided the ideation process. In this section, we examine the challenges of automated loops in HMATs and explore how human-orchestrated teams emerged to address these limitations. We then discuss design considerations for HMAT formation that enable users to effectively orchestrate multiple agents through scalable multi-party communication and progressive team evolution.

\subsection{Breaking the Unproductive Loop: Human-Orchestrated HMATs}
In developing HAT for co-creation, a key consideration has been how to distribute roles between AI agents and humans~\cite{caterina2024customizing, schecter2025how}. Creative workflows that evolve through iterative processes involve two primary roles: the generative role, which generates creative outputs, and the reflective role, which evaluates outputs or provides reflective questions to facilitate further development~\cite{xu2025productive, schecter2025how}. Previous studies have explored various trade-offs in these role distributions—when AI assumes the generative role, diverse idea exploration becomes possible but user agency weakens; when AI takes the reflective role, it induces deeper user reflection but the burden of idea generation remains with humans~\cite{xu2025productive}. 

Our study enabled participants to configure HMAT formations beyond single-agent constraints, particularly by assigning AI agents both generative and reflective roles to automate iterations of the ideation process. However, contrary to their expectations, the automated interactions among agents often devolved into unproductive loops: lacking the capacity to direct and prioritize ideas, the agents repeatedly circled around similar concepts rather than advancing the ideation. Our participants noted that while the evaluations and feedback provided by agents were generally valid, their lack of personal preferences and inability to make value judgments prevented them from giving clear direction. This observation aligns with prior research indicating that, while AI agents can readily surface a wide range of alternatives, they struggle to exercise the subjective value judgments and directional choices that move creative work forward~\cite{gero2023social}. This limitation reflects not only current agent capabilities but also the fundamental nature of creative teaming, where progress depends on subjective value judgments and directional choices in creative work that has no predefined correct answer. It therefore points to the need to ensure that humans hold the steering wheel to provide the direction that creative work inherently requires.

Consequently, most participants ultimately assumed the reflective role themselves to break the unproductive loop, directly evaluating ideas and providing direction. While they actively took on the reflective role in ideation, they continued to explore ways of leveraging multi-agent setups without relying solely on automated ideation. For instance, they operated multiple ideation threads in parallel to secure broader exploration spaces than traditional single HAT, or employed assistant agents that offered alternative reflective perspectives to scaffold human judgment. As they reconfigured how they worked with agents, their role evolved from a narrow reflective role to that of an orchestrator coordinating an entire team. In this role, they synthesized outputs from each thread, set priorities, and decided which ideas to develop next, serving as the central axis that enabled multiple agents to function harmoniously as a team.

Taken together, we suggest forming HMATs in ways that position humans as orchestrators of multiple agents rather than delegating primary control to fully autonomous agent teams. Prior work in human–robot teaming has similarly suggested formations where humans orchestrate multiple agents, acting as managers or supervisors~\cite{gao2012teamwork, abhinav2023survey}. However, in creative work with multiple agents, orchestration involves more than a managerial role. Our findings revealed that users needed to lead the team while simultaneously acting as both managers and active contributors who coordinate the process and set the creative direction. This expanded role, while unlocking the potential of human-orchestrated HMATs, also places a significant burden on users and makes team performance heavily dependent on their capabilities. These tensions surface a new central design question: how should HMATs be designed so that teams remain human-orchestrated while reducing the cognitive and managerial load required for users to fulfill this demanding role? We therefore call for future research that investigates more fine-grained and context-specific human-orchestrated HMAT formations that keep users actively involved while reducing their burden. This agenda extends not only to HMAT formation itself but also to the interaction and system design required to realize such formations. In the following section, we discuss the specific challenges users encounter when orchestrating multiple agents and propose design considerations to address them.

\subsection{Toward Scalable Human Orchestration: Supporting Multi-Party Communications from the User's Perspective}
While participants adopted this human-orchestrated formation to leverage diverse perspectives from multiple agents, managing them simultaneously imposed substantial cognitive load. Our participants noted that this challenge stemmed from a lack of orchestration-specific support in \sysname{}, pointing out that its current interaction and interface design were not optimized for orchestrating multiple agents. This underscores that establishing effective human-orchestrated teams requires more than just identifying the optimal team formation, but also demands the deliberate design of interactions and interfaces that enable users to orchestrate these formations.

Previous HAT research has emphasized the importance of continuous communication and coordination for aligning goals and thoughts among multiple independent entities performing distributed tasks~\cite{zhang2023investigating}. Building on this foundation, HMATs introduce new communication challenges as they require orchestrating interactions among more than two members. Our \sysname{} enabled one-on-one interactions with multiple agents but did not support simultaneous multi-agent interactions, and participants noted this absence as a barrier to real teamwork. For instance, participants struggled to implement situations where all team members participate simultaneously, such as brainstorming sessions, or where they provide direction to the entire team. As participants suggested, future systems should explore interaction methods that allow conducting discussions with multiple agents simultaneously or issuing commands to multiple agents at once. Our findings also revealed that participants directly controlled shared mental models to convey information efficiently without the need for detailed explanations. This finding supports prior research proposing group-level communication strategies, where users treat multiple agents as cohesive groups rather than managing each individually to reduce cognitive burden in multi-agent orchestration~\cite{schömbs2025conversation}. Building on this insight, future work could explore information injection methods beyond direct dialogue when designing HMAT communication systems.

Beyond the challenges of multi-party communication, our findings reveal that participants needed to observe inter-agent interactions to manage their teams. However, they struggled with this monitoring task due to cognitive overload from the sheer volume of agent-to-agent communications, making it challenging to track team dynamics and identify unproductive patterns. Furthermore, requiring agents to communicate in natural language for human comprehension may be inherently inefficient—as previous research has noted, forcing AI to use human-style dialogue can waste time and computational resources when agents could communicate more efficiently through other means~\cite{bucher2024talking, zhang2023investigating}. This presents a fundamental tension between the need for transparency in agent interactions and the practical limitations of human attention and processing capacity. Future research should investigate how to present inter-agent interactions to users in a way that balances transparency with cognitive manageability. These findings suggest opportunities to explore new user interfaces for HMATs that specifically address the challenges of multi-party communication and observing inter-agent interactions while managing multiple agents.

\subsection{Progressive Team Evolution: Human-Aligned HMATs Through Iterative Refinement}
Our findings revealed that participants initially struggled to form teams that functioned as they intended or expected. Despite these early difficulties, we observed participants iteratively developing their teams through trial and error, gradually evolving them into configurations capable of producing ideas that aligned with their personal expectations and ideation contexts. This progressive team development has long been recognized as crucial in human-only teams, with foundational models like Tuckman's team development model (forming, storming, norming, performing) demonstrating how teams evolve through conflict, coordination, and collaboration~\cite{tuckman1965developmental, denise201040, lykourentzou2016team}. This suggests that effective team formation is built over time and interaction, as it involves numerous complex factors that can produce unexpected dynamics and emergent behaviors. Particularly when forming HMATs, it becomes crucial that users themselves take charge of team formation so that the resulting configurations are well aligned with their nuanced goals and working styles. Therefore, we emphasize that HMAT design should focus not on forming perfect teams from the outset, but on empowering users to form and personalize their teams through progressive development.

Prior work on human team formation has proposed tools and strategies that support progressive approaches to team design, enabling people to quickly experience different team configurations and iteratively refine them based on observed outcomes~\cite{lykourentzou2016team, whiting2020parallel}. Building on this line of work, our findings extend progressive team formation to HMATs via \sysname{}, which allows users to form teams with customized AI agents. However, through the process of developing and studying \sysname{}, we found that progressive team growth in HMATs differs in important ways from human team development. Whereas human team formation typically involves recruiting from a bounded pool of candidates and reconfiguring teams at substantial social and organizational cost, HMATs let users instantiate and revise AI agents with comparatively little friction. While this flexibility lets users consider many more plausible team formations, our findings show that adding or reshuffling agents is not reliably associated with increased team performance. In HMATs, progressive improvement thus takes a different form: rather than slowly refining teams drawn from a bounded pool, users tend to rapidly explore, compare, and prune many alternative team formations while continually deciding which agents to instantiate, retain, or retire and how to structure roles within the team. Future research should explore methods for empowering users to progressively form and personalize HMAT formation within such iterative cycles, offering appropriate freedom to explore this enlarged design space without overwhelming them with complexity.

In addition, our findings revealed that participants with practical teamwork experience considered how team members could progressively grow in capability as they performed ideation tasks. However, HMATs required fundamentally different strategies for both team formation development and individual members' growth compared to human teams. Our findings showed that traditional team growth strategies leveraging human psychology—such as inducing competition or fostering senior-junior collaboration—proved ineffective with AI agents, as they lack emotional motivation and social learning capabilities. Rather than emphasizing the capability development of individual team members as in human teams, participants found it more practical to rapidly replace underperforming AI agents with better-configured alternatives. While prior work has proposed similar approaches that employ reinforcement learning and other training techniques to improve the composition of multiple agents for MAS~\cite{lowe2017multi}, HMATs require not only more capable agents but also agents whose roles and behaviors stay aligned with a particular user's goals, values, and expectations for how the team should function. Taken together, these findings point to future work on HMATs that enable users to actively develop both the overall team formation and the capabilities of individual agents over time.

\subsection{Limitations and Future Works}
In this section, we discuss the limitations of our study that could impact the generalization of our findings. 

First, we conducted an exploratory study with 12 participants, which was not a longitudinal investigation. While this approach provided deep insights into HMAT formation, the limited sample size and duration may not capture the full spectrum of strategies that might emerge over the long term. Additionally, we relied on users' subjective judgments rather than measuring actual ideation performance. As a result, we were not able to quantitatively evaluate how the team formations proposed by participants affected team outcomes, such as improvements in idea quality. Although the present study does not aim to identify a single optimal team configuration, the considerations and hypotheses we propose in the discussion require further empirical investigation. Future work should therefore examine how different team formations affect creative outputs, using metrics such as idea quality, novelty, diversity, and feasibility.

Second, our investigation focused exclusively on ideation tasks using iterative divergent-convergent processes. This narrow scope may not generalize to other creative workflows requiring different collaboration patterns. Sequential workflows in software development or design projects might require fundamentally different orchestration strategies. Similarly, creative tasks that require specialized expertise may require different approaches to the distribution of human-agent roles. Further research is needed to understand how HMATs should be formed across diverse creative domains and workflow types.

Third, we examined only single-human multi-agent teams, leaving questions about scenarios with multiple human collaborators. Multi-human HMATs introduce additional complexity in coordination, authority distribution, and conflict resolution that our study did not address. When multiple humans each orchestrate their own agents while collaborating toward shared goals, the interaction patterns are likely to differ substantially from those in single-human scenarios. Future work should investigate these multi-human–multi-agent team formations and develop guidelines for scaling HMATs beyond individual use.

Lastly, our study did not deeply explore the ethical implications of HMATs in creative work. Previous HAT research has documented risks of over-reliance on AI and diminished human autonomy~\cite{duan2025trusting}, and with multiple agents, these risks may compound. In fact, most participants delegated idea generation to AI agents, which could lead to a diminished sense of creative agency and ownership. As HMATs become prevalent, future research should investigate how to preserve human authorship and accountability when forming multi-agent teams, developing guidelines that balance AI assistance with human creative autonomy in team design decisions.

%% file: contents/07conclusion.tex
\section{Conclusion}
In this study, we developed \sysname{}, a technology probe that enables users to form and collaborate with HMATs, allowing us to observe how users specify HMAT formations in practice and surface practical considerations around HMAT formation. Through a three-hour user study with 12 IT design practitioners, in which participants iteratively formed and refined teams across three cycles, we examined how users form HMATs and leverage them in creative ideation processes. We found that while participants initially attempted to let teams operate autonomously, they soon discovered limitations in AI's ability to make value judgments and express clear preferences, which are required for creative work. Consequently, participants adopted a formation where they directly orchestrated agents to break unproductive loops and provide direction. Based on these findings, we emphasize the importance of designing HMATs centered on human orchestration and suggest design considerations that support multi-party communication and progressive team evolution. We hope this research provides insights and a future research agenda for designing human-orchestrated multi-agent teams.

%% file: contents/11appendix.tex
\begin{figure*} [h]
\begin{center}
    \includegraphics[width=\textwidth]{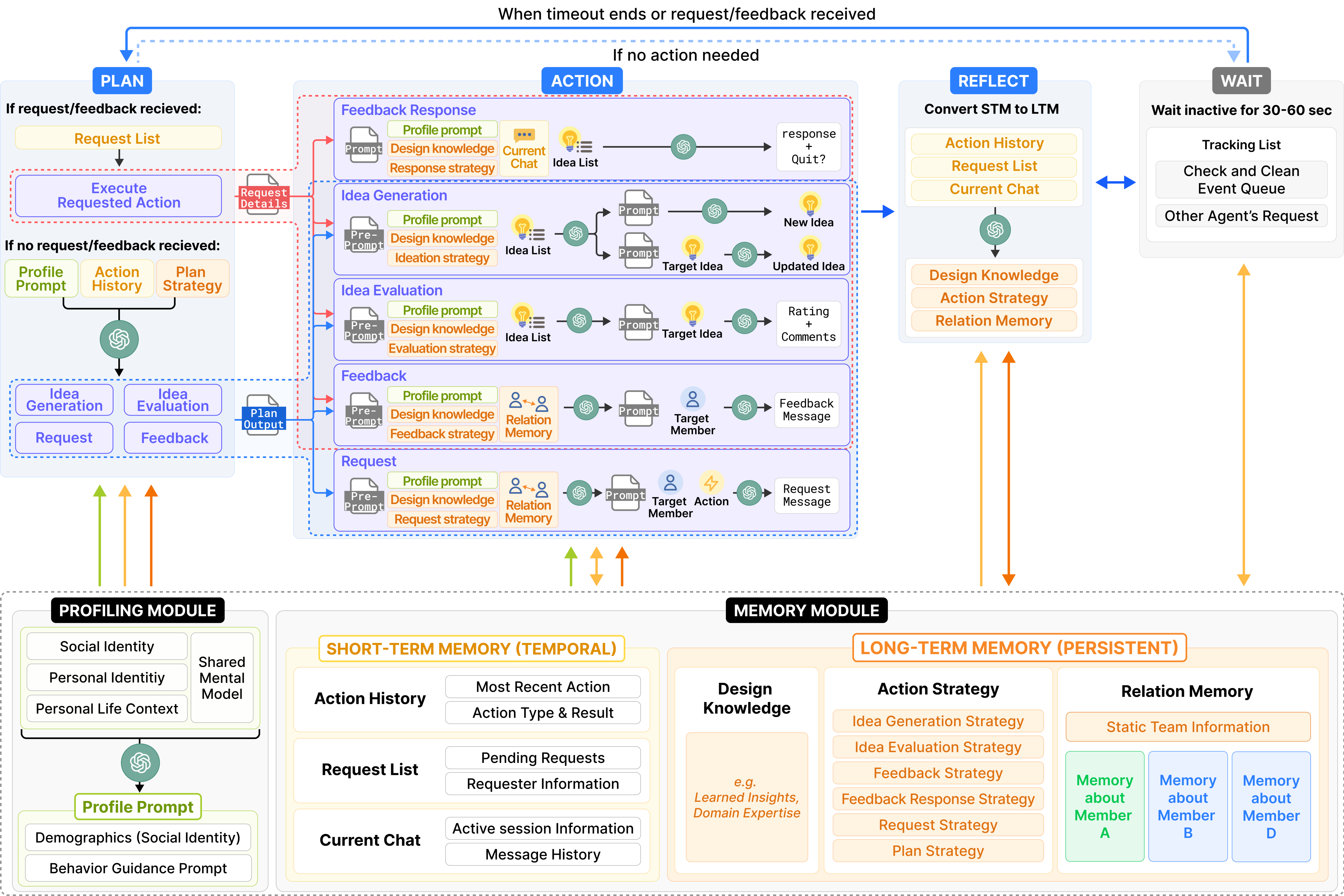}
    \caption[]{Agent architecture of \sysname{}. Agents iterate through Plan → Act → Reflect → Wait states. During the Act state, role-permitted action pipelines are executed. Agent operations are governed by two modules: a Profiling module and a Memory module that separates short-term memory from long-term memory.}
    \label{fig:agent_architecture}
    \vspace{-0.5cm}
\end{center}
\end{figure*}

\section{Detail of System}
\label{sec:system_detail}

\subsection{Design of LLM-Based Agent for HMATs}
\label{sec:agent}

In this section, we describe how we designed AI agents for HMATs in \sysname{}. To enable ideation with multi-agents, we designed agents capable of autonomous interaction while adhering to user customizations (Fig~\ref{fig:agent_architecture}). We adopted the generative agent architecture~\cite{park2023generative, wang2024survey} as our foundation, creating LLM-based agents with a behavioral framework that enables dynamic interaction and environmental responsiveness. This framework operates through four distinct states: (i) Plan, (ii) Act, (iii) Reflect, and (iv) Wait. Users can monitor each agent's state in real-time through the Team Status Tab (Fig.~\ref{fig:crafteam_ideation_UI}.B).

\begin{itemize}
    \item \textbf{Plan State}: In the Plan state, the agent determines its next action. It first checks for any incoming requests from the user or other agents. If a request is present, the agent plans to address it. If there are no pending requests, the agent autonomously selects an action to perform based on its assigned roles.
    \item \textbf{Act State}: In the Act state, the agent executes a single action corresponding to its assigned roles. The possible actions include: Idea Generation, Idea Evaluation, Feedback (and Feedback Response), and Request.
    \item \textbf{Reflect State}: In the Reflect state, the agent processes new information and updates its internal memory. This state is typically triggered after the agent receives an evaluation for an idea it generated or at the conclusion of a feedback session.
    \item \textbf{Wait State}: In the Wait state, the agent remains idle and performs no actions. After a duration of 30-60 seconds, it automatically transitions back to the Plan state. This period serves as a buffer to prepare for new interactions. If the agent receives a direct request or feedback during this time, it immediately transitions to the Act state to respond.
\end{itemize}

To implement this behavioral framework, we designed three core components: a profiling module (Section~\ref{sec:prfiling_module}), a memory module (Section~\ref{sec:memory_module}), and a set of pipelines to execute each action (Section~\ref{sec:action_pipeline}).

\subsubsection{Profiling Module}
\label{sec:prfiling_module}
We designed a profiling module to enable agents to simulate behaviors according to user-defined profiles. Following prior research~\cite{lee2025spectrum, wang2025user}, the profiling module guides the LLMs to embody a character by emphasizing how traits manifest in personal and social contexts rather than merely listing profile attributes. Upon completing team building, the module generates tailored prompts for each agent based on the agent's assigned profile and the team's shared mental model. These prompts translate the profile into behaviorally grounded guidance for the agent, supporting more authentic persona simulation.

\subsubsection{Memory Module}
\label{sec:memory_module}
We designed a memory module to enable agents to remember their prior interactions and reflect on them in subsequent actions. Building on prior research on simulating human-like agents~\cite{wang2025user}, we designed each agent with both Short-Term Memory for recent interaction and Long-Term Memory for enduring information.

Short-Term Memory stores three types of information: (i) five most recent actions to inform subsequent decisions, (ii) a queue of incoming Requests queued for later processing, and (iii) the running transcript of any ongoing Feedback conversation. This memory is populated in real-time and is transient, as its contents are cleared once a request is performed or a feedback session concludes.

Long-Term Memory stores three components: (i) Design Knowledge, (ii) Action Strategies, and (iii) Relationships. Design Knowledge records ideation-relevant facts, constraints, and examples accumulated during the session. Action Strategies specify what the agent may do and how for each role-permitted action (Idea Generation, Idea Evaluation, Feedback, Requests) and the Plan. Given the team-based interaction setting of HMATs, we introduce Relationships, which capture the agent's own working beliefs about other members—who they are and how they connect to the agent (roles and links), salient interaction history with each, and perceived reliability or responsiveness. The agent uses this to decide whom to engage or which action to direct to whom.

\subsubsection{Action Pipelines}
\label{sec:action_pipeline}
To govern the behavior of AI agents in \sysname{}, we designed LLM-based action pipelines that direct agents to perform defined tasks. We implemented two categories of actions: (i) foundational actions, which represent the default capabilities inherent to every agent, such as Plan, Reflection, and Feedback Response; and (ii) role-permitted actions, which are actions specifically assigned to each agent based on their role—such as Idea Generation, Idea Evaluation, Feedback, and Request.

All prompts of LLM-based pipelines consist of both system prompts and user prompts. The system prompt includes the agent profile prompt generated in the profiling module, with additional memory and contextual information assigned to both the system and user prompts depending on the specific action. When a request triggers an action (e.g., Idea Generation, Idea Evaluation, or Feedback), the corresponding request details are consistently appended to the prompt. Following is a detailed description of the pipeline for each action.

\ipstart{Plan Pipeline}
The Plan pipeline determines the following action an AI agent should take. Its inputs include the agent's recent behaviors from short-term memory, as well as design knowledge and relevant action plans from long-term memory. The output requires the agent to either select one of its assigned actions or choose to wait, and to generate a rationale for this decision to support Chain-of-Thought prompting.

\ipstart{Idea Generation Pipeline}
The Idea Generation pipeline enables AI agents to generate ideas. First, through a pre-generation prompt, the agent determines its ideation strategy. The inputs include a list of existing ideas, knowledge from long-term memory, and ideation-related action plans stored in long-term memory. The outputs comprise a decision on whether to create a new idea or develop an existing one, including specification of which idea to develop. Based on this decision, the agent either generates a new idea using the ideation strategy or updates an existing idea by building upon it with the chosen strategy. The idea generation process is implemented using prompts adapted from prior research on co-ideation~\cite{shen2025ideationweb} and employs few-shot prompting.

\ipstart{Idea Evaluation Pipeline}
The Idea Evaluation pipeline enables AI agents to evaluate ideas. First, through a pre-evaluation prompt, the agent determines which idea to evaluate and how to approach the evaluation. The inputs include a list of existing ideas, knowledge from long-term memory, and evaluation-related action plans stored in long-term memory. The outputs comprise the selection of which idea to evaluate. Based on these decisions, the agent evaluates the selected idea and outputs a summary assessment along with scores for each evaluation criterion.

\ipstart{Feedback Pipeline}
The Feedback pipeline enables AI agents to give feedback. First, through a pre-feedback prompt, the agent determines to whom and how to provide feedback. The inputs include a list of existing ideas, relationship information with directly connected team members, knowledge from long-term memory, and feedback-related action plans stored in long-term memory. The outputs comprise the selection of the feedback recipient. Based on these decisions, the agent generates appropriate feedback for the selected recipient. To ensure agents provide constructive and relevant feedback, we incorporated a taxonomy for design idea feedback~\cite{lim2025feedometer} into the prompt to ensure the generation of contextually appropriate feedback. 

\ipstart{Feedback Response Pipeline}
The Feedback Response pipeline enables AI agents to generate responses when receiving feedback. The prompt directs the agent to generate an appropriate response to the received feedback and includes a decision mechanism for determining whether to conclude the feedback session. This allows the agent to terminate the feedback exchange when it judges that further continuation is unnecessary. 

\ipstart{Request Pipeline}
The Request pipeline enables AI agents to make requests. First, through a pre-request prompt, the agent determines to whom and what type of request to send. The inputs include a list of existing ideas, relationship information with directly connected team members, knowledge from long-term memory, and request-related action plans stored in long-term memory. The outputs comprise the selection of the request recipient and the specific action to request. Based on these decisions, the agent generates an appropriate request for the selected recipient.

\input{tables/team_ideation}

\ipstart{Reflection Pipeline}
The Reflection pipeline enables AI agents to update their memory based on past interactions. The agent performs reflection on each evaluation received for its ideas or each feedback session it participates in. Based on these inputs, it (i) extracts new Design Knowledge and adds it to long-term memory, (ii) revises relevant Action Strategies (e.g., when/how to generate, evaluate, or request), and (iii) updates Relationships with the members involved. Upon completing reflection, the agent transitions to the Wait state.

\subsection{Technical Implementation}
\sysname{} is built using Next.js with TypeScript for both frontend and backend development. We employ Upstash Redis as the database for storing user data and team configurations and leverage OpenAI's gpt-4o-2024-08-06 model to power the entire system pipeline. For model parameters, we set the temperature to 0.5 for general system operations to ensure consistent responses, while increasing it to 0.8 for creative tasks, such as ideation, to encourage diverse idea generation.

\section{User Study}
\subsection{Detail of User Study Results}
\label{sec:ideation_detail}
Table~\ref{tab:actors-by-cycle} provides descriptive statistics on how often and in what patterns users and AI agents performed role-permitted actions in each cycle.

%% file: tables/team_ideation.tex
\begin{table*}[t]
\centering
\sffamily
\small
\setlength{\tabcolsep}{6pt}
\begin{tabular}{l l r r r r r r}
\toprule
\textbf{Role} & \textbf{Metric} & \multicolumn{3}{c}{\textbf{Users}} & \multicolumn{3}{c}{\textbf{Agents}} \\
\cmidrule(lr){3-5}\cmidrule(lr){6-8}
 &  & \textbf{Cycle 1} & \textbf{Cycle 2} & \textbf{Cycle 3} & \textbf{Cycle 1} & \textbf{Cycle 2} & \textbf{Cycle 3} \\
\toprule
\multirow{3}{*}{\shortstack[l]{Idea\\Generation}}
 & Count (N)                           & 2 & 3 & 2 & 149 & 143 & 152 \\
 & Per member (M$\pm$SD)              & 0.17$\pm$0.37 & 0.25$\pm$0.60 & 0.17$\pm$0.37  & 3.49$\pm$1.18 & 5.07$\pm$2.99 & 4.54$\pm$2.67 \\
 & Idea length (syll.)                    & 69.50$\pm$3.50 & 79.67$\pm$73.07 & 117.50$\pm$62.50 & 157.39$\pm$81.74 & 176.17$\pm$75.84 & 168.67$\pm$82.36 \\
\midrule
\multirow{6}{*}{\shortstack[l]{Idea\\Evaluation}}
 & Count (N)                           & 36 & 42 & 40 & 118 & 80 & 112 \\
 & Per member (M$\pm$SD)              & 4.00$\pm$2.87 & 4.20$\pm$2.23 & 4.00$\pm$2.14 & 2.73$\pm$1.16 & 3.20$\pm$1.34 & 3.62$\pm$2.06 \\
 & Comment length (syll.)                 & 46.31$\pm$31.72  & 43.79$\pm$34.21 & 44.80$\pm$32.36 & 290.77$\pm$46.30 & 305.94$\pm$47.33 & 296.17$\pm$42.92 \\
 & rating: Novelty (M)                     & 3.25   & 4.26   & 4.78   & 5.50   & 5.39   & 5.38   \\
 & rating: Completeness (M)              & 3.06   & 3.83   & 4.38   & 5.19   & 5.29   & 5.31   \\
 & rating: Quality (M)                     & 3.06   & 4.07   & 4.38   & 5.40   & 5.42   & 5.42   \\
\midrule
\multirow{4}{*}{Feedback}
 & Session Count (N)                   &  18  & 30  &  38  & 120  & 63  & 89    \\
 & Per member (M$\pm$SD)               & 1.50$\pm$1.38 & 2.50$\pm$1.71 & 3.17$\pm$2.27 & 2.73$\pm$1.16 & 2.10$\pm$1.28 & 2.07$\pm$1.47 \\
 & Message length (syll.)                 & 42.92$\pm$29.37  & 32.42$\pm$23.84 & 36.57$\pm$23.39 & 75.21$\pm$30.14  & 75.48$\pm$34.12 & 74.69$\pm$32.11 \\
 & Turns (M$\pm$SD)                    & 2.83$\pm$0.96 & 3.17$\pm$1.21 & 2.58$\pm$1.14 & 6.46$\pm$1.99 & 5.56$\pm$1.97 & 5.65$\pm$2.10 \\
\midrule
\multirow{6}{*}{Requests}
 & Count (N)                           & 32 & 16 & 24 & 3 & 21 & 10 \\
 & Per member (M$\pm$SD)               & 2.67$\pm$2.17 & 1.33$\pm$1.43 & 2.00$\pm$2.00 & 0.12$\pm$0.32 & 0.91$\pm$1.50 & 0.38$\pm$0.79 \\
 & Message length (syll.)                 & 36.69$\pm$25.10 & 30.88$\pm$19.58 & 27.83$\pm$13.59 & 161.33$\pm$11.03 & 189.48$\pm$24.96 & 181.10$\pm$20.62 \\
 & Type: Generation (N)                & 22 & 11 & 16 & 0 & 0 & 1 \\
 & Type: Evaluation (N)                & 7 & 3 & 5 & 0 & 9 & 3 \\
 & Type: Feedback (N)                  & 3 & 2 & 3 & 3 & 12 & 6 \\
\bottomrule
\end{tabular}
\vspace{0.2cm}
\caption[Comprehensive table showing role-based ideation activities across three cycles, comparing Users and Agents performance. For Idea Generation: Users generated 2-3 ideas per cycle (0.17-0.25 per member average) while Agents produced 143-152 ideas (3.49-5.07 per member), with idea lengths ranging from 69-117 syllables for Users and 157-176 for Agents. For Idea Evaluation: Users made 36-42 evaluations (4.00-4.20 per member) compared to Agents' 80-118 evaluations (2.73-3.62 per member), with Users providing shorter comments (43-46 syllables) than Agents (290-305 syllables), and both groups rating ideas on Novelty, Completeness, and Quality scales. For Feedback: Users conducted 18-38 sessions (1.50-3.17 per member) while Agents held 63-120 sessions (2.07-2.73 per member), with conversation turns averaging 2.58-3.17 for Users and 5.56-6.46 for Agents. For Requests: Users made 16-32 requests (1.33-2.67 per member) primarily for Generation tasks, while Agents made 3-21 requests (0.12-0.91 per member) mainly for Evaluation and Feedback tasks.]{Descriptive statistics of role-permitted actions in team-based ideation session: frequency and details of Idea Generation, Idea Evaluation, Feedback, and Requests by user and agents across three cycles.}
\label{tab:actors-by-cycle}
\end{table*}

%% file: contents/10bibliography.bib
@inproceedings{hutchinson2003technology,
author = {Hutchinson, Hilary and Mackay, Wendy and Westerlund, Bo and Bederson, Benjamin B. and Druin, Allison and Plaisant, Catherine and Beaudouin-Lafon, Michel and Conversy, St\'{e}phane and Evans, Helen and Hansen, Heiko and Roussel, Nicolas and Eiderb\"{a}ck, Bj\"{o}rn},
title = {Technology probes: inspiring design for and with families},
year = {2003},
isbn = {1581136307},
publisher = {Association for Computing Machinery},
address = {New York, NY, USA},
url = {https://doi.org/10.1145/642611.642616},
doi = {10.1145/642611.642616},
booktitle = {Proceedings of the SIGCHI Conference on Human Factors in Computing Systems},
pages = {17–24},
numpages = {8},
location = {Ft. Lauderdale, Florida, USA},
series = {CHI '03}
}

@ARTICLE{iftikhar2024human,
  author={Iftikhar, Rehan and Chiu, Yi-Te and Khan, Mohammad Saud and Caudwell, Catherine},
  journal={IEEE Transactions on Engineering Management}, 
  title={Human–Agent Team Dynamics: A Review and Future Research Opportunities}, 
  year={2024},
  volume={71},
  number={},
  pages={10139-10154},
  doi={10.1109/TEM.2023.3331369}}

@inproceedings{shen2025ideationweb,
author = {Shen, Hanshu and Shen, Lyukesheng and Wu, Wenqi and Zhang, Kejun},
title = {IdeationWeb: Tracking the Evolution of Design Ideas in Human-AI Co-Creation},
year = {2025},
isbn = {9798400713941},
publisher = {Association for Computing Machinery},
address = {New York, NY, USA},
url = {https://doi.org/10.1145/3706598.3713375},
doi = {10.1145/3706598.3713375},
booktitle = {Proceedings of the 2025 CHI Conference on Human Factors in Computing Systems},
articleno = {146},
numpages = {19},
location = {
},
series = {CHI '25}
}

@inproceedings{park2023generative,
author = {Park, Joon Sung and O'Brien, Joseph and Cai, Carrie Jun and Morris, Meredith Ringel and Liang, Percy and Bernstein, Michael S.},
title = {Generative Agents: Interactive Simulacra of Human Behavior},
year = {2023},
isbn = {9798400701320},
publisher = {Association for Computing Machinery},
address = {New York, NY, USA},
url = {https://doi.org/10.1145/3586183.3606763},
doi = {10.1145/3586183.3606763},
booktitle = {Proceedings of the 36th Annual ACM Symposium on User Interface Software and Technology},
articleno = {2},
numpages = {22},
location = {San Francisco, CA, USA},
series = {UIST '23}
}

@inproceedings{xu2025productive,
author = {Xu, Xiaotong (Tone) and Konnova, Arina and Gao, Bianca and Peng, Cindy and Vo, Dave and Dow, Steven P.},
title = {Productive vs. Reflective: How Different Ways of Integrating AI into Design Workflows Affect Cognition and Motivation},
year = {2025},
isbn = {9798400713941},
publisher = {Association for Computing Machinery},
address = {New York, NY, USA},
url = {https://doi.org/10.1145/3706598.3713649},
doi = {10.1145/3706598.3713649},
booktitle = {Proceedings of the 2025 CHI Conference on Human Factors in Computing Systems},
articleno = {24},
numpages = {15},
location = {
},
series = {CHI '25}
}

@inproceedings{kim2025amuse,
author = {Kim, Yewon and Lee, Sung-Ju and Donahue, Chris},
title = {Amuse: Human-AI Collaborative Songwriting with Multimodal Inspirations},
year = {2025},
isbn = {9798400713941},
publisher = {Association for Computing Machinery},
address = {New York, NY, USA},
url = {https://doi.org/10.1145/3706598.3713818},
doi = {10.1145/3706598.3713818},
booktitle = {Proceedings of the 2025 CHI Conference on Human Factors in Computing Systems},
articleno = {187},
numpages = {28},
location = {
},
series = {CHI '25}
}

@inproceedings{han2025choreocraft,
author = {Han, Hyunyoung and Jung, Kyungeun and Yoon, Sang Ho},
title = {ChoreoCraft: In-situ Crafting of Choreography in Virtual Reality through Creativity Support Tool},
year = {2025},
isbn = {9798400713941},
publisher = {Association for Computing Machinery},
address = {New York, NY, USA},
url = {https://doi.org/10.1145/3706598.3714220},
doi = {10.1145/3706598.3714220},
booktitle = {Proceedings of the 2025 CHI Conference on Human Factors in Computing Systems},
articleno = {1059},
numpages = {21},
location = {
},
series = {CHI '25}
}

@inproceedings{choi2024creativeconnect,
author = {Choi, DaEun and Hong, Sumin and Park, Jeongeon and Chung, John Joon Young and Kim, Juho},
title = {CreativeConnect: Supporting Reference Recombination for Graphic Design Ideation with Generative AI},
year = {2024},
isbn = {9798400703300},
publisher = {Association for Computing Machinery},
address = {New York, NY, USA},
url = {https://doi.org/10.1145/3613904.3642794},
doi = {10.1145/3613904.3642794},
booktitle = {Proceedings of the 2024 CHI Conference on Human Factors in Computing Systems},
articleno = {1055},
numpages = {25},
location = {Honolulu, HI, USA},
series = {CHI '24}
}

@inproceedings{naik2025designing,
author = {Naik, Suchismita and Toombs, Austin L. and Snellinger, Amanda, Ph.D. and Saponas, Scott and Hall, Amanda K},
title = {Designing with Multi-Agent Generative AI: Insights from Industry Early Adopters},
year = {2025},
isbn = {9798400714856},
publisher = {Association for Computing Machinery},
address = {New York, NY, USA},
url = {https://doi.org/10.1145/3715336.3735823},
doi = {10.1145/3715336.3735823},
booktitle = {Proceedings of the 2025 ACM Designing Interactive Systems Conference},
pages = {1961–1972},
numpages = {12},
location = {
},
series = {DIS '25}
}

@INPROCEEDINGS{ding2023designGPT,
  author={Ding, Shiying and Chen, Xinyi and Fang, Yan and Liu, Wenrui and Qiu, Yiwu and Chai, Chunlei},
  booktitle={2023 16th International Symposium on Computational Intelligence and Design (ISCID)}, 
  title={DesignGPT: Multi-Agent Collaboration in Design}, 
  year={2023},
  volume={},
  number={},
  pages={204-208},
  doi={10.1109/ISCID59865.2023.00056}}

@inproceedings{lim2024co-creating,
author = {Lim, Hyunseung and Cho, Ji Yong and Kim, Taewan and Park, Jeongeon and Shin, Hyungyu and Choi, Seulgi and Park, Sunghyun and Lee, Kyungjae and Kim, Juho and Lee, Moontae and Hong, Hwajung},
title = {Co-Creating Question-and-Answer Style Articles with Large Language Models for Research Promotion},
year = {2024},
isbn = {9798400705830},
publisher = {Association for Computing Machinery},
address = {New York, NY, USA},
url = {https://doi.org/10.1145/3643834.3660705},
doi = {10.1145/3643834.3660705},
booktitle = {Proceedings of the 2024 ACM Designing Interactive Systems Conference},
pages = {975–994},
numpages = {20},
location = {Copenhagen, Denmark},
series = {DIS '24}
}

@article{james2019team,
author = {James C. Walliser and Ewart J. de Visser and Eva Wiese and Tyler H. Shaw},
title ={Team Structure and Team Building Improve Human–Machine Teaming With Autonomous Agents},
journal = {Journal of Cognitive Engineering and Decision Making},
volume = {13},
number = {4},
pages = {258-278},
year = {2019},
doi = {10.1177/1555343419867563},
URL = { 
        https://doi.org/10.1177/1555343419867563
}
}

@inproceedings{chen2024agentverse,
 author = {Chen, Weize and Su, Yusheng and Zuo, Jingwei and Yang, Cheng and Yuan, Chenfei and Chan, Chi-Min and Yu, Heyang and Lu, Yaxi and Hung, Yi-Hsin and Qian, Chen and Qin, Yujia and Cong, Xin and Xie, Ruobing and Liu, Zhiyuan and Sun, Maosong and Zhou, Jie},
 booktitle = {International Conference on Representation Learning},
 editor = {B. Kim and Y. Yue and S. Chaudhuri and K. Fragkiadaki and M. Khan and Y. Sun},
 pages = {20094--20136},
 title = {AgentVerse: Facilitating Multi-Agent Collaboration and Exploring Emergent Behaviors},
 url = {https://proceedings.iclr.cc/paper\_files/paper/2024/file/578e65cdee35d00c708d4c64bce32971-Paper-Conference.pdf},
 volume = {2024},
 year = {2024}
}

@Article{wang2024survey,
author={Wang, Lei
and Ma, Chen
and Feng, Xueyang
and Zhang, Zeyu
and Yang, Hao
and Zhang, Jingsen
and Chen, Zhiyuan
and Tang, Jiakai
and Chen, Xu
and Lin, Yankai
and Zhao, Wayne Xin
and Wei, Zhewei
and Wen, Jirong},
title={A survey on large language model based autonomous agents},
journal={Frontiers of Computer Science},
year={2024},
month={Mar},
day={22},
volume={18},
number={6},
pages={186345},
issn={2095-2236},
doi={10.1007/s11704-024-40231-1},
url={https://doi.org/10.1007/s11704-024-40231-1}
}

@article{xu2024adaptation,
    author = {Xu, Zeda and Hong, Chloe Soohwa and Soria Zurita, Nicolás F. and Gyory, Joshua T. and Stump, Gary and Nolte, Hannah and Cagan, Jonathan and McComb, Christopher},
    title = {Adaptation Through Communication: Assessing Human–Artificial Intelligence Partnership for the Design of Complex Engineering Systems},
    journal = {Journal of Mechanical Design},
    volume = {146},
    number = {8},
    pages = {081401},
    year = {2024},
    month = {02},
    issn = {1050-0472},
    doi = {10.1115/1.4064490},
    url = {https://doi.org/10.1115/1.4064490}
}

@inproceedings{khan2025beyond,
author = {Khan, Abidullah and Shokrizadeh, Atefeh and Cheng, Jinghui},
title = {Beyond Automation: How Designers Perceive AI as a Creative Partner in the Divergent Thinking Stages of UI/UX Design},
year = {2025},
isbn = {9798400713941},
publisher = {Association for Computing Machinery},
address = {New York, NY, USA},
url = {https://doi.org/10.1145/3706598.3713500},
doi = {10.1145/3706598.3713500},
booktitle = {Proceedings of the 2025 CHI Conference on Human Factors in Computing Systems},
articleno = {1105},
numpages = {12},
location = {
},
series = {CHI '25}
}

@article{rezwana2023designing,
author = {Rezwana, Jeba and Maher, Mary Lou},
title = {Designing Creative AI Partners with COFI: A Framework for Modeling Interaction in Human-AI Co-Creative Systems},
year = {2023},
issue_date = {October 2023},
publisher = {Association for Computing Machinery},
address = {New York, NY, USA},
volume = {30},
number = {5},
issn = {1073-0516},
url = {https://doi.org/10.1145/3519026},
doi = {10.1145/3519026},
journal = {ACM Trans. Comput.-Hum. Interact.},
month = sep,
articleno = {67},
numpages = {28}
}

@article{thomas2022Human,
author = {Thomas O’Neill and Nathan McNeese and Amy Barron and Beau Schelble},
title ={Human-Autonomy Teaming: A Review and Analysis of the Empirical Literature},
journal = {Human Factors},
volume = {64},
number = {5},
pages = {904-938},
year = {2022},
doi = {10.1177/0018720820960865},
    note ={PMID: 33092417},
URL = { 
    https://doi.org/10.1177/0018720820960865
}
}

@ARTICLE{kaelin2024developing,
AUTHOR={Kaelin, Vera C.  and Tewari, Maitreyee  and Benouar, Sara  and Lindgren, Helena },
TITLE={Developing teamwork: transitioning between stages in human-agent collaboration},
JOURNAL={Frontiers in Computer Science},
VOLUME={Volume 6 - 2024},
YEAR={2024},
URL={https://www.frontiersin.org/journals/computer-science/articles/10.3389/fcomp.2024.1455903},
DOI={10.3389/fcomp.2024.1455903},
ISSN={2624-9898}}

@article{zhang2023investigating,
author = {Zhang, Rui and Duan, Wen and Flathmann, Christopher and McNeese, Nathan and Freeman, Guo and Williams, Alyssa},
title = {Investigating AI Teammate Communication Strategies and Their Impact in Human-AI Teams for Effective Teamwork},
year = {2023},
issue_date = {October 2023},
publisher = {Association for Computing Machinery},
address = {New York, NY, USA},
volume = {7},
number = {CSCW2},
url = {https://doi.org/10.1145/3610072},
doi = {10.1145/3610072},
journal = {Proc. ACM Hum.-Comput. Interact.},
month = oct,
articleno = {281},
numpages = {31}
}

@inproceedings{guo2024large,
  title     = {Large Language Model Based Multi-agents: A Survey of Progress and Challenges},
  author    = {Guo, Taicheng and Chen, Xiuying and Wang, Yaqi and Chang, Ruidi and Pei, Shichao and Chawla, Nitesh V. and Wiest, Olaf and Zhang, Xiangliang},
  booktitle = {Proceedings of the Thirty-Third International Joint Conference on
               Artificial Intelligence, {IJCAI-24}},
  publisher = {International Joint Conferences on Artificial Intelligence Organization},
  editor    = {Kate Larson},
  pages     = {8048--8057},
  year      = {2024},
  month     = {8},
  note      = {Survey Track},
  doi       = {10.24963/ijcai.2024/890},
  url       = {https://doi.org/10.24963/ijcai.2024/890},
}

@ARTICLE{dorst2006design,
  author={Dorst, Kees},
  journal={Design Issues}, 
  title={Design Problems and Design Paradoxes}, 
  year={2006},
  volume={22},
  number={3},
  pages={4-17},
  doi={10.1162/desi.2006.22.3.4}}

@article{he2025llm,
author = {He, Junda and Treude, Christoph and Lo, David},
title = {LLM-Based Multi-Agent Systems for Software Engineering: Literature Review, Vision, and the Road Ahead},
year = {2025},
issue_date = {June 2025},
publisher = {Association for Computing Machinery},
address = {New York, NY, USA},
volume = {34},
number = {5},
issn = {1049-331X},
url = {https://doi.org/10.1145/3712003},
doi = {10.1145/3712003},
journal = {ACM Trans. Softw. Eng. Methodol.},
month = May,
articleno = {124},
numpages = {30}
}

@INPROCEEDINGS{nguyen2025agilecoder,
  author={Nguyen, Minh Huynh and Phan Chau, Thang and Nguyen, Phong X. and Bui, Nghi D. Q.},
  booktitle={2025 IEEE/ACM Second International Conference on AI Foundation Models and Software Engineering (Forge)}, 
  title={AgileCoder: Dynamic Collaborative Agents for Software Development based on Agile Methodology}, 
  year={2025},
  volume={},
  number={},
  pages={156-167},
  doi={10.1109/Forge66646.2025.00026}}

@misc{jin2024mare,
      title={MARE: Multi-Agents Collaboration Framework for Requirements Engineering}, 
      author={Dongming Jin and Zhi Jin and Xiaohong Chen and Chunhui Wang},
      year={2024},
      eprint={2405.03256},
      archivePrefix={arXiv},
      primaryClass={cs.SE},
      url={https://arxiv.org/abs/2405.03256}, 
}

@misc{lin2025creativity,
      title={Creativity in LLM-based Multi-Agent Systems: A Survey}, 
      author={Yi-Cheng Lin and Kang-Chieh Chen and Zhe-Yan Li and Tzu-Heng Wu and Tzu-Hsuan Wu and Kuan-Yu Chen and Hung-yi Lee and Yun-Nung Chen},
      year={2025},
      eprint={2505.21116},
      archivePrefix={arXiv},
      primaryClass={cs.HC},
      url={https://arxiv.org/abs/2505.21116}, 
}

@ARTICLE{dorri2018multi,
  author={Dorri, Ali and Kanhere, Salil S. and Jurdak, Raja},
  journal={IEEE Access}, 
  title={Multi-Agent Systems: A Survey}, 
  year={2018},
  volume={6},
  number={},
  pages={28573-28593},
  doi={10.1109/ACCESS.2018.2831228}}

@ARTICLE{rezaee2015average,
  author={Rezaee, Hamed and Abdollahi, Farzaneh},
  journal={IEEE Transactions on Automatic Control}, 
  title={Average Consensus Over High-Order Multiagent Systems}, 
  year={2015},
  volume={60},
  number={11},
  pages={3047-3052},
  doi={10.1109/TAC.2015.2408576}}

@article{wan2024investigating,
author = {Wan, Qian and Hu, Siying and Zhang, Yu and Wang, Piaohong and Wen, Bo and Lu, Zhicong},
title = {"It Felt Like Having a Second Mind": Investigating Human-AI Co-creativity in Prewriting with Large Language Models},
year = {2024},
issue_date = {April 2024},
publisher = {Association for Computing Machinery},
address = {New York, NY, USA},
volume = {8},
number = {CSCW1},
url = {https://doi.org/10.1145/3637361},
doi = {10.1145/3637361},
journal = {Proc. ACM Hum.-Comput. Interact.},
month = apr,
articleno = {84},
numpages = {26}
}

@inproceedings{chen2024hollmwood,
    title = "{H}o{LLM}wood: Unleashing the Creativity of Large Language Models in Screenwriting via Role Playing",
    author = "Chen, Jing  and
      Zhu, Xinyu  and
      Yang, Cheng  and
      Shi, Chufan  and
      Xi, Yadong  and
      Zhang, Yuxiang  and
      Wang, Junjie  and
      Pu, Jiashu  and
      Feng, Tian  and
      Yang, Yujiu  and
      Zhang, Rongsheng",
    editor = "Al-Onaizan, Yaser  and
      Bansal, Mohit  and
      Chen, Yun-Nung",
    booktitle = "Findings of the Association for Computational Linguistics: EMNLP 2024",
    month = nov,
    year = "2024",
    address = "Miami, Florida, USA",
    publisher = "Association for Computational Linguistics",
    url = "https://aclanthology.org/2024.findings-emnlp.474/",
    doi = "10.18653/v1/2024.findings-emnlp.474",
    pages = "8075--8121"
}

@misc{wan2025using,
      title={Using Generative AI Personas Increases Collective Diversity in Human Ideation}, 
      author={Yun Wan and Yoram M Kalman},
      year={2025},
      eprint={2504.13868},
      archivePrefix={arXiv},
      primaryClass={cs.HC},
      url={https://arxiv.org/abs/2504.13868}, 
}

@inproceedings{tian2024macgyver,
    title = "{M}ac{G}yver: Are Large Language Models Creative Problem Solvers?",
    author = "Tian, Yufei  and
      Ravichander, Abhilasha  and
      Qin, Lianhui  and
      Le Bras, Ronan  and
      Marjieh, Raja  and
      Peng, Nanyun  and
      Choi, Yejin  and
      Griffiths, Thomas  and
      Brahman, Faeze",
    editor = "Duh, Kevin  and
      Gomez, Helena  and
      Bethard, Steven",
    booktitle = "Proceedings of the 2024 Conference of the North American Chapter of the Association for Computational Linguistics: Human Language Technologies (Volume 1: Long Papers)",
    month = jun,
    year = "2024",
    address = "Mexico City, Mexico",
    publisher = "Association for Computational Linguistics",
    url = "https://aclanthology.org/2024.naacl-long.297/",
    doi = "10.18653/v1/2024.naacl-long.297",
    pages = "5303--5324"
}

@article{abhinav2023survey,
title = {A survey of multi-agent Human–Robot Interaction systems},
journal = {Robotics and Autonomous Systems},
volume = {161},
pages = {104335},
year = {2023},
issn = {0921-8890},
doi = {https://doi.org/10.1016/j.robot.2022.104335},
url = {https://www.sciencedirect.com/science/article/pii/S092188902200224X},
author = {Abhinav Dahiya and Alexander M. Aroyo and Kerstin Dautenhahn and Stephen L. Smith}
}

@inproceedings{li2023camel,
 author = {Li, Guohao and Hammoud, Hasan and Itani, Hani and Khizbullin, Dmitrii and Ghanem, Bernard},
 booktitle = {Advances in Neural Information Processing Systems},
 editor = {A. Oh and T. Naumann and A. Globerson and K. Saenko and M. Hardt and S. Levine},
 pages = {51991--52008},
 publisher = {Curran Associates, Inc.},
 title = {CAMEL: Communicative Agents for "Mind" Exploration of Large Language Model Society},
 url = {https://proceedings.neurips.cc/paper\_files/paper/2023/file/a3621ee907def47c1b952ade25c67698-Paper-Conference.pdf},
 volume = {36},
 year = {2023}
}

@article{robert2023role,
author = {Robert W. Andrews and J. Mason Lilly and Divya Srivastava and Karen M. Feigh},
title = {The role of shared mental models in human-AI teams: a theoretical review},
journal = {Theoretical Issues in Ergonomics Science},
volume = {24},
number = {2},
pages = {129--175},
year = {2023},
publisher = {Taylor \& Francis},
doi = {10.1080/1463922X.2022.2061080},
URL = { 
        https://doi.org/10.1080/1463922X.2022.2061080
}
}

@article{bansa2019beyond, 
title={Beyond Accuracy: The Role of Mental Models in Human-AI Team Performance}, volume={7}, url={https://ojs.aaai.org/index.php/HCOMP/article/view/5285}, DOI={10.1609/hcomp.v7i1.5285}, abstractNote={&lt;p&gt;Decisions made by human-AI teams (&lt;em&gt;e.g&lt;/em&gt;., AI-advised humans) are increasingly common in high-stakes domains such as healthcare, criminal justice, and finance. Achieving high &lt;em&gt;team&lt;/em&gt; performance depends on more than just the accuracy of the AI system: Since the human and the AI may have different expertise, the highest team performance is often reached when they both know how and when to complement one another. We focus on a factor that is crucial to supporting such complementary: the human’s mental model of the AI capabilities, specifically the AI system’s &lt;em&gt;error boundary&lt;/em&gt; (&lt;em&gt;i.e.&lt;/em&gt; knowing “When does the AI err?”). Awareness of this lets the human decide when to accept or override the AI’s recommendation. We highlight two key properties of an AI’s error boundary, &lt;em&gt;parsimony&lt;/em&gt; and &lt;em&gt;stochasticity&lt;/em&gt;, and a property of the task, &lt;em&gt;dimensionality&lt;/em&gt;. We show experimentally how these properties affect humans’ mental models of AI capabilities and the resulting team performance. We connect our evaluations to related work and propose goals, beyond accuracy, that merit consideration during model selection and optimization to improve overall human-AI team performance.&lt;/p&gt;}, number={1}, journal={Proceedings of the AAAI Conference on Human Computation and Crowdsourcing}, author={Bansal, Gagan and Nushi, Besmira and Kamar, Ece and Lasecki, Walter S. and Weld, Daniel S. and Horvitz, Eric}, year={2019}, month={Oct.}, pages={2-11} }

@article{zhang2021ideal,
author = {Zhang, Rui and McNeese, Nathan J. and Freeman, Guo and Musick, Geoff},
title = {"An Ideal Human": Expectations of AI Teammates in Human-AI Teaming},
year = {2021},
issue_date = {December 2020},
publisher = {Association for Computing Machinery},
address = {New York, NY, USA},
volume = {4},
number = {CSCW3},
url = {https://doi.org/10.1145/3432945},
doi = {10.1145/3432945},
journal = {Proc. ACM Hum.-Comput. Interact.},
month = jan,
articleno = {246},
numpages = {25}
}

@inproceedings{liang2019implicit,
author = {Liang, Claire and Proft, Julia and Andersen, Erik and Knepper, Ross A.},
title = {Implicit Communication of Actionable Information in Human-AI teams},
year = {2019},
isbn = {9781450359702},
publisher = {Association for Computing Machinery},
address = {New York, NY, USA},
url = {https://doi.org/10.1145/3290605.3300325},
doi = {10.1145/3290605.3300325},
booktitle = {Proceedings of the 2019 CHI Conference on Human Factors in Computing Systems},
pages = {1–13},
numpages = {13},
location = {Glasgow, Scotland Uk},
series = {CHI '19}
}

@inproceedings{duan2025trusting,
author = {Duan, Wen and Flathmann, Christopher and McNeese, Nathan and Scalia, Matthew J and Zhang, Ruihao and Gorman, Jamie and Freeman, Guo and Zhou, Shiwen and Hauptman, Allyson Ivy and Yin, Xiaoyun},
title = {Trusting Autonomous Teammates in Human-AI Teams - A Literature Review},
year = {2025},
isbn = {9798400713941},
publisher = {Association for Computing Machinery},
address = {New York, NY, USA},
url = {https://doi.org/10.1145/3706598.3713527},
doi = {10.1145/3706598.3713527},
booktitle = {Proceedings of the 2025 CHI Conference on Human Factors in Computing Systems},
articleno = {1102},
numpages = {23},
location = {
},
series = {CHI '25}
}

@article{nathan2021team,
author = {Nathan J. McNeese and Mustafa Demir and Nancy J. Cooke and Manrong She},
title ={Team Situation Awareness and Conflict: A Study of Human–Machine Teaming},
journal = {Journal of Cognitive Engineering and Decision Making},
volume = {15},
number = {2-3},
pages = {83-96},
year = {2021},
doi = {10.1177/15553434211017354},
URL = { 
        https://doi.org/10.1177/15553434211017354
}
}

@article{schelble2022lets,
author = {Schelble, Beau G. and Flathmann, Christopher and McNeese, Nathan J. and Freeman, Guo and Mallick, Rohit},
title = {Let's Think Together! Assessing Shared Mental Models, Performance, and Trust in Human-Agent Teams},
year = {2022},
issue_date = {January 2022},
publisher = {Association for Computing Machinery},
address = {New York, NY, USA},
volume = {6},
number = {GROUP},
url = {https://doi.org/10.1145/3492832},
doi = {10.1145/3492832},
journal = {Proc. ACM Hum.-Comput. Interact.},
month = jan,
articleno = {13},
numpages = {29}
}

@article{jan2024ai,
title = {AI-teaming: Redefining collaboration in the digital era},
journal = {Current Opinion in Psychology},
volume = {58},
pages = {101837},
year = {2024},
issn = {2352-250X},
doi = {https://doi.org/10.1016/j.copsyc.2024.101837},
url = {https://www.sciencedirect.com/science/article/pii/S2352250X24000502},
author = {Jan B. Schmutz and Neal Outland and Sophie Kerstan and Eleni Georganta and Anna-Sophie Ulfert}
}

@article{duan2024understanding,
author = {Duan, Wen and Zhou, Shiwen and Scalia, Matthew J and Yin, Xiaoyun and Weng, Nan and Zhang, Ruihao and Freeman, Guo and McNeese, Nathan and Gorman, Jamie and Tolston, Michael},
title = {Understanding the Evolvement of Trust Over Time within Human-AI Teams},
year = {2024},
issue_date = {November 2024},
publisher = {Association for Computing Machinery},
address = {New York, NY, USA},
volume = {8},
number = {CSCW2},
url = {https://doi.org/10.1145/3687060},
doi = {10.1145/3687060},
journal = {Proc. ACM Hum.-Comput. Interact.},
month = nov,
articleno = {521},
numpages = {31}
}

@inproceedings{lee2025spectrum,
    title = "{SP}e{C}trum: A Grounded Framework for Multidimensional Identity Representation in {LLM}-Based Agent",
    author = "Lee, Keyeun  and
      Kim, Seo Hyeong  and
      Lee, Seolhee  and
      Eun, Jinsu  and
      Ko, Yena  and
      Jeon, Hayeon  and
      Kim, Esther Hehsun  and
      Cho, Seonghye  and
      Yang, Soeun  and
      Kim, Eun-mee  and
      Lim, Hajin",
    editor = "Chiruzzo, Luis  and
      Ritter, Alan  and
      Wang, Lu",
    booktitle = "Proceedings of the 2025 Conference of the Nations of the Americas Chapter of the Association for Computational Linguistics: Human Language Technologies (Volume 1: Long Papers)",
    month = apr,
    year = "2025",
    address = "Albuquerque, New Mexico",
    publisher = "Association for Computational Linguistics",
    url = "https://aclanthology.org/2025.naacl-long.356/",
    doi = "10.18653/v1/2025.naacl-long.356",
    pages = "6971--6991",
    ISBN = "979-8-89176-189-6"
}

@article{wang2025user,
author = {Wang, Lei and Zhang, Jingsen and Yang, Hao and Chen, Zhi-Yuan and Tang, Jiakai and Zhang, Zeyu and Chen, Xu and Lin, Yankai and Sun, Hao and Song, Ruihua and Zhao, Xin and Xu, Jun and Dou, Zhicheng and Wang, Jun and Wen, Ji-Rong},
title = {User Behavior Simulation with Large Language Model-based Agents},
year = {2025},
issue_date = {March 2025},
publisher = {Association for Computing Machinery},
address = {New York, NY, USA},
volume = {43},
number = {2},
issn = {1046-8188},
url = {https://doi.org/10.1145/3708985},
doi = {10.1145/3708985},
journal = {ACM Trans. Inf. Syst.},
month = jan,
articleno = {55},
numpages = {37}
}

@article{virginia2006thematic,
author = { Virginia   Braun  and  Victoria   Clarke },
title = {Using thematic analysis in psychology},
journal = {Qualitative Research in Psychology},
volume = {3},
number = {2},
pages = {77-101},
year  = {2006},
publisher = {Routledge},
doi = {10.1191/1478088706qp063oa},
URL = { 
        https://www.tandfonline.com/doi/abs/10.1191/1478088706qp063o
}
}

@article{mathieu2019embracing,
   author = {Mathieu, John E. and Gallagher, Peter T. and Domingo, Monique A. and Klock, Elizabeth A.},
   title = {Embracing Complexity: Reviewing the Past Decade of Team Effectiveness Research}, 
   journal= {Annual Review of Organizational Psychology and Organizational Behavior},
   year = {2019},
   volume = {6},
   number = {Volume 6, 2019},
   pages = {17-46},
   doi = {https://doi.org/10.1146/annurev-orgpsych-012218-015106},
   url = {https://www.annualreviews.org/content/journals/10.1146/annurev-orgpsych-012218-015106},
   publisher = {Annual Reviews},
   issn = {2327-0616},
   type = {Journal Article}
  }

@article{steve2006enhancing,
author = {Steve W.J. Kozlowski and Daniel R. Ilgen},
title ={Enhancing the Effectiveness of Work Groups and Teams},
journal = {Psychological Science in the Public Interest},
volume = {7},
number = {3},
pages = {77-124},
year = {2006},
doi = {10.1111/j.1529-1006.2006.00030.x},
    note ={PMID: 26158912},
URL = { 
        https://doi.org/10.1111/j.1529-1006.2006.00030.x
}
}

@proceedings{takai2017towards,
    author = {Takai, Shun and Esterman, Marcos},
    title = {Towards a Better Design Team Formation: A Review of Team Effectiveness Models and Possible Measurements of Design-Team Inputs, Processes, and Outputs},
    volume = {Volume 3: 19th International Conference on Advanced Vehicle Technologies; 14th International Conference on Design Education; 10th Frontiers in Biomedical Devices},
    series = {International Design Engineering Technical Conferences and Computers and Information in Engineering Conference},
    pages = {V003T04A018},
    year = {2017},
    month = {08},
    doi = {10.1115/DETC2017-68091},
    url = {https://doi.org/10.1115/DETC2017-68091}
}

@article{perry2003social,
author = {Perry-Smith, Jill E. and Shalley, Christina E.},
title = {The Social Side of Creativity: A Static and Dynamic Social Network Perspective},
journal = {Academy of Management Review},
volume = {28},
number = {1},
pages = {89-106},
year = {2003},
doi = {10.5465/amr.2003.8925236},
URL = { 
        https://doi.org/10.5465/amr.2003.8925236
}
}

@article{roger2003virtuality,
title = {Virtuality, communication, and new product team creativity: a social network perspective},
journal = {Journal of Engineering and Technology Management},
volume = {20},
number = {1},
pages = {69-92},
year = {2003},
note = {Special Issue on Research Issues in Knowledge Management and Virtual Collaboration in New Product Development},
issn = {0923-4748},
doi = {https://doi.org/10.1016/S0923-4748(03)00005-5},
url = {https://www.sciencedirect.com/science/article/pii/S0923474803000055},
author = {Roger Th.A.J Leenders and Jo M.L {van Engelen} and Jan Kratzer}
}

@article{thomas2023human,
title = {Human-autonomy Teaming: Need for a guiding team-based framework?},
journal = {Computers in Human Behavior},
volume = {146},
pages = {107762},
year = {2023},
issn = {0747-5632},
doi = {https://doi.org/10.1016/j.chb.2023.107762},
url = {https://www.sciencedirect.com/science/article/pii/S0747563223001139},
author = {Thomas A. O'Neill and Christopher Flathmann and Nathan J. McNeese and Eduardo Salas}
}

@article{jo2001improving,
author = {Jo M. L. van Engelen and Derk Jan Kiewiet and Pieter Terlouw},
title = {Improving Performance of Product Development Teams through Managing Polarity},
journal = {International Studies of Management \& Organization},
volume = {31},
number = {1},
pages = {46--63},
year = {2001},
publisher = {Routledge},
doi = {10.1080/00208825.2001.11656807},
URL = { 
        https://doi.org/10.1080/00208825.2001.11656807
}
}

@article{stefan2007increasing,
author = {Stefan Wuchty  and Benjamin F. Jones  and Brian Uzzi },
title = {The Increasing Dominance of Teams in Production of Knowledge},
journal = {Science},
volume = {316},
number = {5827},
pages = {1036-1039},
year = {2007},
doi = {10.1126/science.1136099},
URL = {https://www.science.org/doi/abs/10.1126/science.1136099}}

@article{aritzeta2007belbin,
author = {Aritzeta, Aitor and Swailes, Stephen and Senior, Barbara},
title = {Belbin's Team Role Model: Development, Validity and Applications for Team Building},
journal = {Journal of Management Studies},
volume = {44},
number = {1},
pages = {96-118},
doi = {https://doi.org/10.1111/j.1467-6486.2007.00666.x},
url = {https://onlinelibrary.wiley.com/doi/abs/10.1111/j.1467-6486.2007.00666.x},
year = {2007}
}

@article{tuckman1965developmental,
  author = {Tuckman, Bruce W.},
  title = {Developmental sequence in small groups},
  year = {1965},
  volume = {63},
  number = {6},
  doi = {https://doi.org/10.1037/h0022100},
  publisher = {American Psychological Association},
  issn = {1939-1455(Electronic);0033-2909(Print)},
}

@article{mathieu2017century,
  author = {Mathieu, John E. and Hollenbeck, John R. and van Knippenberg, Daan and Ilgen, Daniel R.},
  title = {A century of work teams in the Journal of Applied Psychology},
  year = {2017},
  volume = {102},
  number = {3},
  doi = {10.1037/apl0000128},
  publisher = {American Psychological Association},
  isbn = {978-1-4338-9042-0},
  issn = {1939-1854(Electronic);0021-9010(Print)}
}

@article{denise201040,
author = {Denise A. Bonebright},
title = {40 years of storming: a historical review of Tuckman's model of small group development},
journal = {Human Resource Development International},
volume = {13},
number = {1},
pages = {111--120},
year = {2010},
publisher = {Routledge},
doi = {10.1080/13678861003589099},
URL = { 
        https://doi.org/10.1080/13678861003589099
}
}

@Article{toh2016creativity,
author={Toh, Christine A.
and Miller, Scarlett R.},
title={Creativity in design teams: the influence of personality traits and risk attitudes on creative concept selection},
journal={Research in Engineering Design},
year={2016},
month={Jan},
day={01},
volume={27},
number={1},
pages={73-89},
issn={1435-6066},
doi={10.1007/s00163-015-0207-y},
url={https://doi.org/10.1007/s00163-015-0207-y}
}

@article{jan2008social,
title = {The social structure of leadership and creativity in engineering design teams: An empirical analysis},
journal = {Journal of Engineering and Technology Management},
volume = {25},
number = {4},
pages = {269-286},
year = {2008},
issn = {0923-4748},
doi = {https://doi.org/10.1016/j.jengtecman.2008.10.004},
url = {https://www.sciencedirect.com/science/article/pii/S0923474808000441},
author = {Jan Kratzer and Roger Th.A.J. Leenders and Jo M.L. {Van Engelen}}
}

@inproceedings{yu2025systematic,
  author    = {Yu, Zixiao and Cui, Tingru and Luo, Chen and Tan, Dilang},
  title     = {A Systematic Literature Review on Human-Agent Teaming with Insights into Multi-Agent Interactions},
  booktitle = {Proceedings of the Pacific Asia Conference on Information Systems (PACIS) 2025},
  year      = {2025},
  volume    = {20},
  url       = {https://aisel.aisnet.org/pacis2025/hci/hci/20},
  note      = {Track 5: Human Computer Interaction, Paper Number: PACIS2025-1614}
}

@inproceedings{song2025the,
title={The More, The Stronger? Investigating How Multi-Agent {AI} Shapes Human Opinions},
author={Tianqi Song and Yugin Tan and Zicheng Zhu and Maojia Song and Feng Yibin and Yi-Chieh Lee},
booktitle={ICLR 2025 Workshop on Human-AI Coevolution},
year={2025},
url={https://openreview.net/forum?id=6zlttMWe4G}
}

@article{nathan2018teaming,
author = {Nathan J. McNeese and Mustafa Demir and Nancy J. Cooke and Christopher Myers},
title ={Teaming With a Synthetic Teammate: Insights into Human-Autonomy Teaming},
journal = {Human Factors},
volume = {60},
number = {2},
pages = {262-273},
year = {2018},
doi = {10.1177/0018720817743223},
    note ={PMID: 29185818},
URL = { 
    https://doi.org/10.1177/0018720817743223
}
}

@inproceedings{caterina2024customizing,
  author    = {Caterina Moruzzi and Solange Margarido},
  title     = {Customizing the Balance between User and System Agency in Human-{AI} Co-Creative Processes},
  booktitle = {Proceedings of the 15th International Conference on Computational Creativity (ICCC'24)},
  editor    = {Kazjon Grace and Maria Teresa Llano and Pedro Martins and Maria M. Hedblom},
  address   = {J{\"o}nk{\"o}ping, Sweden},
  pages     = {108--117},
  publisher = {Association for Computational Creativity},
  year      = {2024},
  url       = {https://computationalcreativity.net/iccc24/papers/ICCC24_paper_15.pdf}
}

@inproceedings{schecter2025how,
author = {Schecter, Aaron and Richardson, Benjamin},
title = {How the Role of Generative AI Shapes Perceptions of Value in Human-AI Collaborative Work},
year = {2025},
isbn = {9798400713941},
publisher = {Association for Computing Machinery},
address = {New York, NY, USA},
url = {https://doi.org/10.1145/3706598.3713946},
doi = {10.1145/3706598.3713946},
booktitle = {Proceedings of the 2025 CHI Conference on Human Factors in Computing Systems},
articleno = {530},
numpages = {15},
location = {
},
series = {CHI '25}
}

@article{lim2025feedometer,
title = {Feed-o-meter: Investigating AI-generated mentee personas as interactive agents for scaffolding design feedback practice},
journal = {International Journal of Human-Computer Studies},
pages = {103687},
year = {2025},
issn = {1071-5819},
doi = {https://doi.org/10.1016/j.ijhcs.2025.103687},
url = {https://www.sciencedirect.com/science/article/pii/S1071581925002447},
author = {Hyunseung Lim and Dasom Choi and DaEun Choi and Sooyohn Nam and Hwajung Hong}
}

@misc{park2025choicemates,
      title={ChoiceMates: Supporting Unfamiliar Online Decision-Making with Multi-Agent Conversational Interactions}, 
      author={Jeongeon Park and Bryan Min and Kihoon Son and Jean Y. Song and Xiaojuan Ma and Juho Kim},
      year={2025},
      eprint={2310.01331},
      archivePrefix={arXiv},
      primaryClass={cs.HC},
      url={https://arxiv.org/abs/2310.01331}, 
}

@inproceedings{nomura2024towards,
author = {Nomura, Moeka and Ito, Takayuki and Ding, Shiyao},
title = {Towards Collaborative Brain-storming among Humans and AI Agents: An Implementation of the IBIS-based Brainstorming Support System with Multiple AI Agents},
year = {2024},
isbn = {9798400705540},
publisher = {Association for Computing Machinery},
address = {New York, NY, USA},
url = {https://doi.org/10.1145/3643562.3672609},
doi = {10.1145/3643562.3672609},
booktitle = {Proceedings of the ACM Collective Intelligence Conference},
pages = {1–9},
numpages = {9},
location = {Boston, MA, USA},
series = {CI '24}
}

@inproceedings{ghosh2025yes,
author = {Ghosh, Pratik and Rintel, Sean},
title = {YES AND: A Generative AI Multi-Agent Framework for Enhancing Diversity of Thought in Individual Ideation for Problem-Solving Through Confidence-Based Agent Turn-Taking},
year = {2025},
isbn = {9798400713958},
publisher = {Association for Computing Machinery},
address = {New York, NY, USA},
url = {https://doi.org/10.1145/3706599.3720142},
doi = {10.1145/3706599.3720142},
booktitle = {Proceedings of the Extended Abstracts of the CHI Conference on Human Factors in Computing Systems},
articleno = {607},
numpages = {13},
location = {
},
series = {CHI EA '25}
}

@article{julie1995workload,
author = {Julie M. Urban and Clint A. Bowers and Susan D. Monday and Ben B. Morgan Jr.},
title = {Workload, Team Structure, and Communication in Team Performance},
journal = {Military Psychology},
volume = {7},
number = {2},
pages = {123--139},
year = {1995},
publisher = {Routledge},
doi = {10.1207/s15327876mp0702\_6},
URL = {  
    https://doi.org/10.1207/s15327876mp0702\_6
}}

@article{bucher2024talking,
author = {Bucher, Andreas and Dolata, Mateusz and Eckhardt, Sven and Staehelin, Dario and Schwabe, Gerhard},
title = {Talking to Multi-Party Conversational Agents in Advisory Services: Command-based vs. Conversational Interactions},
year = {2024},
issue_date = {January 2024},
publisher = {Association for Computing Machinery},
address = {New York, NY, USA},
volume = {8},
number = {GROUP},
url = {https://doi.org/10.1145/3633072},
doi = {10.1145/3633072},
journal = {Proc. ACM Hum.-Comput. Interact.},
month = feb,
articleno = {7},
numpages = {25}
}

@inproceedings{gero2023social,
author = {Gero, Katy Ilonka and Long, Tao and Chilton, Lydia B},
title = {Social Dynamics of AI Support in Creative Writing},
year = {2023},
isbn = {9781450394215},
publisher = {Association for Computing Machinery},
address = {New York, NY, USA},
url = {https://doi.org/10.1145/3544548.3580782},
doi = {10.1145/3544548.3580782},
booktitle = {Proceedings of the 2023 CHI Conference on Human Factors in Computing Systems},
articleno = {245},
numpages = {15},
location = {Hamburg, Germany},
series = {CHI '23}
}

@misc{schömbs2025conversation,
      title={From Conversation to Orchestration: HCI Challenges and Opportunities in Interactive Multi-Agentic Systems}, 
      author={Sarah Schömbs and Yan Zhang and Jorge Goncalves and Wafa Johal},
      year={2025},
      eprint={2506.20091},
      archivePrefix={arXiv},
      primaryClass={cs.HC},
      url={https://arxiv.org/abs/2506.20091}, 
}

@inproceedings{he2024ai,
author = {He, Jessica and Houde, Stephanie and Gonzalez, Gabriel E. and Silva Moran, Dar\'{\i}o Andr\'{e}s and Ross, Steven I. and Muller, Michael and Weisz, Justin D.},
title = {AI and the Future of Collaborative Work: Group Ideation with an LLM in a Virtual Canvas},
year = {2024},
isbn = {9798400710179},
publisher = {Association for Computing Machinery},
address = {New York, NY, USA},
url = {https://doi.org/10.1145/3663384.3663398},
doi = {10.1145/3663384.3663398},
booktitle = {Proceedings of the 3rd Annual Meeting of the Symposium on Human-Computer Interaction for Work},
articleno = {9},
numpages = {14},
location = {Newcastle upon Tyne, United Kingdom},
series = {CHIWORK '24}
}

@inproceedings{lowe2017multi,
author = {Lowe, Ryan and Wu, Yi and Tamar, Aviv and Harb, Jean and Abbeel, Pieter and Mordatch, Igor},
title = {Multi-agent actor-critic for mixed cooperative-competitive environments},
year = {2017},
isbn = {9781510860964},
publisher = {Curran Associates Inc.},
address = {Red Hook, NY, USA},
booktitle = {Proceedings of the 31st International Conference on Neural Information Processing Systems},
pages = {6382–6393},
numpages = {12},
location = {Long Beach, California, USA},
series = {NIPS'17}
}

@InProceedings{schulte2016design,
author="Schulte, Axel
and Donath, Diana
and Lange, Douglas S.",
editor="Harris, Don",
title="Design Patterns for Human-Cognitive Agent Teaming",
booktitle="Engineering Psychology and Cognitive Ergonomics",
year="2016",
publisher="Springer International Publishing",
address="Cham",
pages="231--243",
isbn="978-3-319-40030-3"
}

@inproceedings{figueroa2019automatic,
author = {Figueroa, Hugo and Costaguta, Rosanna and Menini, Mar\'{\i}a de los \'{A}ngeles and Missio, Daniela},
title = {An Automatic Identification of Team Roles in Forums},
year = {2019},
isbn = {9781450371766},
publisher = {Association for Computing Machinery},
address = {New York, NY, USA},
url = {https://doi.org/10.1145/3335595.3335615},
doi = {10.1145/3335595.3335615},
booktitle = {Proceedings of the XX International Conference on Human Computer Interaction},
articleno = {24},
numpages = {2},
location = {Donostia, Gipuzkoa, Spain},
series = {Interacci\'{o}n '19}
}

@INPROCEEDINGS{paruchuri2010effect,
  author={Paruchuri, Praveen and Varakantham, Pradeep and Sycara, Katia and Scerri, Paul},
  booktitle={2010 IEEE/WIC/ACM International Conference on Web Intelligence and Intelligent Agent Technology}, 
  title={Effect of Human Biases on Human-Agent Teams}, 
  year={2010},
  volume={2},
  number={},
  pages={327-334},
  doi={10.1109/WI-IAT.2010.104}}

@article{dow2011parallel,
author = {Dow, Steven P. and Glassco, Alana and Kass, Jonathan and Schwarz, Melissa and Schwartz, Daniel L. and Klemmer, Scott R.},
title = {Parallel prototyping leads to better design results, more divergence, and increased self-efficacy},
year = {2011},
issue_date = {December 2010},
publisher = {Association for Computing Machinery},
address = {New York, NY, USA},
volume = {17},
number = {4},
issn = {1073-0516},
url = {https://doi.org/10.1145/1879831.1879836},
doi = {10.1145/1879831.1879836},
journal = {ACM Trans. Comput.-Hum. Interact.},
month = dec,
articleno = {18},
numpages = {24}
}

@inproceedings{jung2013engaging,
author = {Jung, Malte F. and Lee, Jin Joo and DePalma, Nick and Adalgeirsson, Sigurdur O. and Hinds, Pamela J. and Breazeal, Cynthia},
title = {Engaging robots: easing complex human-robot teamwork using backchanneling},
year = {2013},
isbn = {9781450313315},
publisher = {Association for Computing Machinery},
address = {New York, NY, USA},
url = {https://doi.org/10.1145/2441776.2441954},
doi = {10.1145/2441776.2441954},
booktitle = {Proceedings of the 2013 Conference on Computer Supported Cooperative Work},
pages = {1555–1566},
numpages = {12},
location = {San Antonio, Texas, USA},
series = {CSCW '13}
}

@article{sebo2020robots,
author = {Sebo, Sarah and Stoll, Brett and Scassellati, Brian and Jung, Malte F.},
title = {Robots in Groups and Teams: A Literature Review},
year = {2020},
issue_date = {October 2020},
publisher = {Association for Computing Machinery},
address = {New York, NY, USA},
volume = {4},
number = {CSCW2},
url = {https://doi.org/10.1145/3415247},
doi = {10.1145/3415247},
journal = {Proc. ACM Hum.-Comput. Interact.},
month = oct,
articleno = {176},
numpages = {36}
}

@article{matthias2017framework,
author = {Matthias Scheutz and Scott A. DeLoach and Julie A. Adams},
title ={A Framework for Developing and Using Shared Mental Models in Human-Agent Teams},
journal = {Journal of Cognitive Engineering and Decision Making},
volume = {11},
number = {3},
pages = {203-224},
year = {2017},
doi = {10.1177/1555343416682891},
URL = { 
        https://doi.org/10.1177/1555343416682891
}
}

@inproceedings{dietz2017human,
author = {Dietz, Griffin and E, Jane L. and Washington, Peter and Kim, Lawrence H. and Follmer, Sean},
title = {Human Perception of Swarm Robot Motion},
year = {2017},
isbn = {9781450346566},
publisher = {Association for Computing Machinery},
address = {New York, NY, USA},
url = {https://doi.org/10.1145/3027063.3053220},
doi = {10.1145/3027063.3053220},
booktitle = {Proceedings of the 2017 CHI Conference Extended Abstracts on Human Factors in Computing Systems},
pages = {2520–2527},
numpages = {8},
location = {Denver, Colorado, USA},
series = {CHI EA '17}
}

@inproceedings{kim2020bot,
author = {Kim, Soomin and Eun, Jinsu and Oh, Changhoon and Suh, Bongwon and Lee, Joonhwan},
title = {Bot in the Bunch: Facilitating Group Chat Discussion by Improving Efficiency and Participation with a Chatbot},
year = {2020},
isbn = {9781450367080},
publisher = {Association for Computing Machinery},
address = {New York, NY, USA},
url = {https://doi.org/10.1145/3313831.3376785},
doi = {10.1145/3313831.3376785},
booktitle = {Proceedings of the 2020 CHI Conference on Human Factors in Computing Systems},
pages = {1–13},
numpages = {13},
location = {Honolulu, HI, USA},
series = {CHI '20}
}

@ARTICLE{vig2006multi,
  author={Vig, L. and Adams, J.A.},
  journal={IEEE Transactions on Robotics}, 
  title={Multi-robot coalition formation}, 
  year={2006},
  volume={22},
  number={4},
  pages={637-649},
  doi={10.1109/TRO.2006.878948}}

@inproceedings{lykourentzou2016team,
author = {Lykourentzou, Ioanna and Wang, Shannon and Kraut, Robert E. and Dow, Steven P.},
title = {Team Dating: A Self-Organized Team Formation Strategy for Collaborative Crowdsourcing},
year = {2016},
isbn = {9781450340823},
publisher = {Association for Computing Machinery},
address = {New York, NY, USA},
url = {https://doi.org/10.1145/2851581.2892421},
doi = {10.1145/2851581.2892421},
booktitle = {Proceedings of the 2016 CHI Conference Extended Abstracts on Human Factors in Computing Systems},
pages = {1243–1249},
numpages = {7},
location = {San Jose, California, USA},
series = {CHI EA '16}
}

@inproceedings{lappas2009finding,
author = {Lappas, Theodoros and Liu, Kun and Terzi, Evimaria},
title = {Finding a team of experts in social networks},
year = {2009},
isbn = {9781605584959},
publisher = {Association for Computing Machinery},
address = {New York, NY, USA},
url = {https://doi.org/10.1145/1557019.1557074},
doi = {10.1145/1557019.1557074},
booktitle = {Proceedings of the 15th ACM SIGKDD International Conference on Knowledge Discovery and Data Mining},
pages = {467–476},
numpages = {10},
location = {Paris, France},
series = {KDD '09}
}

@article{wi2009team,
title = {A team formation model based on knowledge and collaboration},
journal = {Expert Systems with Applications},
volume = {36},
number = {5},
pages = {9121-9134},
year = {2009},
issn = {0957-4174},
doi = {https://doi.org/10.1016/j.eswa.2008.12.031},
url = {https://www.sciencedirect.com/science/article/pii/S095741740800907X},
author = {Hyeongon Wi and Seungjin Oh and Jungtae Mun and Mooyoung Jung}
}

@inproceedings{anagnotopoulos2012online,
author = {Anagnostopoulos, Aris and Becchetti, Luca and Castillo, Carlos and Gionis, Aristides and Leonardi, Stefano},
title = {Online team formation in social networks},
year = {2012},
isbn = {9781450312295},
publisher = {Association for Computing Machinery},
address = {New York, NY, USA},
url = {https://doi.org/10.1145/2187836.2187950},
doi = {10.1145/2187836.2187950},
booktitle = {Proceedings of the 21st International Conference on World Wide Web},
pages = {839–848},
numpages = {10},
location = {Lyon, France},
series = {WWW '12}
}

@inproceedings{gomzzara2019who,
author = {G\'{o}mez-Zar\'{a}, Diego and Paras, Matthew and Twyman, Marlon and Lane, Jacqueline N. and DeChurch, Leslie A. and Contractor, Noshir S.},
title = {Who Would You Like to Work With?},
year = {2019},
isbn = {9781450359702},
publisher = {Association for Computing Machinery},
address = {New York, NY, USA},
url = {https://doi.org/10.1145/3290605.3300889},
doi = {10.1145/3290605.3300889},
booktitle = {Proceedings of the 2019 CHI Conference on Human Factors in Computing Systems},
pages = {1–15},
numpages = {15},
location = {Glasgow, Scotland Uk},
series = {CHI '19}
}

@article{whiting2020parallel,
author = {Whiting, Mark E. and Gao, Irena and Xing, Michelle and Diarrassouba, N'godjigui Junior and Nguyen, Tonya and Bernstein, Michael S.},
title = {Parallel Worlds: Repeated Initializations of the Same Team to Improve Team Viability},
year = {2020},
issue_date = {May 2020},
publisher = {Association for Computing Machinery},
address = {New York, NY, USA},
volume = {4},
number = {CSCW1},
url = {https://doi.org/10.1145/3392877},
doi = {10.1145/3392877},
journal = {Proc. ACM Hum.-Comput. Interact.},
month = may,
articleno = {67},
numpages = {22}
}

@inproceedings{gao2012teamwork,
author = {Gao, Fei and Cummings, Missy L. and Bertuccelli, Luca F.},
title = {Teamwork in controlling multiple robots},
year = {2012},
isbn = {9781450310635},
publisher = {Association for Computing Machinery},
address = {New York, NY, USA},
url = {https://doi.org/10.1145/2157689.2157703},
doi = {10.1145/2157689.2157703},
booktitle = {Proceedings of the Seventh Annual ACM/IEEE International Conference on Human-Robot Interaction},
pages = {81–88},
numpages = {8},
location = {Boston, Massachusetts, USA},
series = {HRI '12}
}
